\documentclass[
twocolumn,
showpacs,
nofootinbib
]{revtex4-1}
\usepackage{amsthm}
\theoremstyle{plain}
\newtheorem{theorem}{Theorem}
\newtheorem*{theorem*}{Theorem}

\newtheorem*{definition*}{Definition}
\newtheorem{lemma}[theorem]{Lemma}
\newtheorem*{lemma*}{Lemma}

\newtheorem{remark}[theorem]{Remark}

\usepackage{braket}
\usepackage{delarray}
\usepackage{here}

\usepackage{amsmath}
\usepackage{txfonts}

\usepackage{bm}
\usepackage{color}
\usepackage{ulem}

\newcommand{\ba}{\begin{array}}
\newcommand{\ea}{\end{array}}
\newcommand{\bmat}{\left(\begin{array}}
\newcommand{\emat}{\end{array}\right)}

\newcommand{\be}{\begin{eqnarray}}
\newcommand{\ee}{\end{eqnarray}}

\newcommand{\RG}{\mathcal R}

\begin{document}
\title{Existence and Absence of Spin-Glass Phases on Migdal--Kadanoff Hierarchical Lattices: A Complete Classification}

\author{Manaka Okuyama$^{1,2}$\footnote{corresponding author: manaka.okuyama.d2@tohoku.ac.jp}}
\author{Masayuki Ohzeki$^{1,3,4,5}$}
\affiliation{$^1$Graduate School of Information Sciences, Tohoku University, Sendai 980-8579, Japan}
\affiliation{$^2$School of Computing, Institute of Science Tokyo, Tokyo 152-8551, Japan}
\affiliation{$^3$Department of Physics, Institute of Science Tokyo, Tokyo 152-8551, Japan}
\affiliation{$^4$Research and Education Institute for Semiconductors and Informatics, Kumamoto University, Kumamoto 860-0862, Japan}
\affiliation{$^5$Sigma-i Co., Ltd., Tokyo 108-0075, Japan} 

\begin{abstract}

We consider the Ising spin-glass model on
Migdal--Kadanoff (MK) hierarchical lattices,
characterized by the fractal dimension
$d=1+\log s/\log b$, where $s$ and $b$ denote
the number of parallel branches and the number
of bonds in series within each branch,
respectively.
For independent Gaussian couplings, we rigorously
establish a necessary and sufficient condition
for the existence of finite-temperature spin-glass
order for arbitrary integers $b,s\ge2$.
We prove that a finite-temperature spin-glass
phase exists whenever
\[
\frac{b}{\sqrt{s}}
\left(
\frac{b\sqrt{s(s-1)}}{bs-1}
\right)^{s-1}<1.
\]
In particular, this condition holds for all
three-dimensional MK hierarchical lattices
($d=3$, or equivalently $s=b^2$),
regardless of the choice of $b$.
Conversely, we prove the absence of
finite-temperature spin-glass order for
fractal dimensions
$d<1+\log17/\log6\approx2.58$,
as well as in the large-$b$ limit for any
fixed $d<3$.
In the intermediate regime $1+\log17/\log6\le d<3$, we provide a finite-step procedure to determine the presence or absence of spin-glass order, which depends on $b$ and $s$.
Our results provide a complete characterization
of finite-temperature spin-glass order on
MK hierarchical lattices.

%
\end{abstract}

\date{\today}
\maketitle
\tableofcontents

\section{Introduction}
While the rigorous analysis of mean-field spin-glass
models has made remarkable progress~\cite{parisi1980sequence,guerra2003broken,talagrand2006parisi}, the mathematical
understanding of finite-dimensional spin glasses
remains limited~\cite{newman1996non,newman2000nature,newman2023proof,chatterjee2023spin,newman2026ground,nishimori2025temperature}.
It is widely believed that a finite-temperature
spin-glass phase exists in three dimensions
but not in two dimensions~\cite{mcmillan1984domain,mcmillan1984domain2,bray1984lower,bhatt1988numerical,hartmann2001lower,houdayer2004low,singh1986critical,kawashima1997finite,lukic2004critical}, although neither
prediction has been rigorously established.
Whether replica symmetry breaking occurs
in three-dimensional spin glasses also remains
an open problem.

The Ising spin-glass model on Migdal--Kadanoff (MK)
hierarchical lattices~\cite{migdal1975phase,kadanoff1976notes}, in which each unit consists
of $s$ parallel branches of $b$ bonds in series,
provides a framework where the renormalization-group
(RG) transformation can be performed exactly~\cite{griffiths1982spin,kaufman1981exactly,kaufman1984spin}.
The resulting RG equations coincide with those
obtained from the MK approximation for
finite-dimensional spin glasses, and the model
has been extensively studied~\cite{young1976real,southern1977real,jayaprakash1977spin,mckay1982spin,bray1987chaotic,curado1988spin,nifle1992new,moore1998evidence,curado1999ground,drossel2000link,amoruso2003scalings,sasaki2003temperature,bouchaud2003energy,jorg2012nature,angelini2015spin,demirtacs2015lower}.
Spin glasses on MK hierarchical lattices
are generally believed to be described by
the droplet picture, without replica symmetry
breaking.
For relatively small values of $b$ and $s$,
the lower critical dimension has been estimated
to be approximately $2.52$~\cite{demirtacs2015lower}, whereas scaling
analyses suggest that it approaches $d=3$
in the large-$b$ limit~\cite{boettcher2005stiffness}.
Nevertheless, rigorous results for spin glasses
on MK hierarchical lattices remain limited.
In particular, the existence of finite-temperature
spin-glass order and the absence of replica symmetry
breaking have been rigorously established only
at sufficiently high fractal dimensions~\cite{gardner1984spin,collet1984spin,kamphorst1986mixed}.
More recently, for symmetric bimodal couplings,
the absence of finite-temperature spin-glass order
has been proved for $(b,s)=(5,6)$ ($d\approx2.11$)
and $(b,s)=(6,8)$ ($d\approx2.16$), as well as
in the large-$b$ limit for $d<3$~\cite{okuyama2026absence}.
However, rigorous proofs of the existence
of finite-temperature spin-glass order
at low fractal dimensions
for finite values of $b$ and $s$
have remained unavailable.

In this work, we consider independent Gaussian
couplings and establish a necessary and sufficient
condition for the existence of finite-temperature
spin-glass order on MK hierarchical lattices
for arbitrary integers $b,s\ge2$.
In particular, we prove that a finite-temperature
spin-glass phase exists whenever
\[
\frac{b}{\sqrt{s}}
\left(
\frac{b\sqrt{s(s-1)}}{bs-1}
\right)^{s-1}<1,
\]
which holds for all $d\ge3$.
Conversely, we prove the absence of
finite-temperature spin-glass order for
$d<1+\log17/\log6\approx2.58$, as well as
in the large-$b$ limit for any fixed $d<3$.
In the intermediate regime
$1+\log17/\log6\le d<3$,
the existence of spin-glass order depends
on the specific values of $b$ and $s$.
We develop a finite-step procedure to determine
the presence or absence of spin-glass order
whenever the RG growth rate differs from
its marginal value.
These results establish a rigorous classification of finite-temperature spin-glass phases on MK hierarchical lattices.

Our proof begins with a rigorous analysis of
the zero-temperature renormalization-group (RG)
equations.
We show that the effective couplings grow
without bound at high fractal
dimensions, whereas they decay to zero
at low fractal dimensions.
Our zero-temperature analysis is inspired
in part by the rigorous RG approach
developed in Ref.~\cite{bovier1991stability}
for a hierarchical random-interface model.

The absence of finite-temperature spin-glass
order follows relatively directly from
the zero-temperature results, together
with a comparison between the finite-
and zero-temperature RG flows.
In contrast, establishing the existence of
finite-temperature spin-glass order requires
a Peierls-type argument adapted to
MK hierarchical lattices.
A key ingredient is a reference spin configuration
that need not be a ground state at finite
temperature but becomes one in the
zero-temperature limit.
Our approach is closely related to the
recursive argument of Collet and
Eckmann~\cite{collet1984spin}, but employs
an explicit reference spin configuration
to control thermal fluctuations.

%

The remainder of this paper is organized as follows.
In Sec.~II, we define the model and present our main results. 
In Sec.~III, we establish preliminary results used in the proofs of the main theorems.
In Sec.~IV, we prove the absence of
finite-temperature spin-glass order.
In Sec.~V, we establish a finite-step procedure
that can, in principle, determine the presence
or absence of finite-temperature spin-glass order,
provided that $\Lambda(b,s)\neq1$.
In Sec.~VI, we prove the existence of
finite-temperature spin-glass order.
Finally, Sec.~VII summarizes our results
and discusses open problems.

%
%

\section{Model and results}

We consider the Ising spin-glass model on the
Migdal--Kadanoff (MK) hierarchical lattice specified
by two integers $b\ge 2$ and $s\ge 2$.
Starting from a single bond at generation $n=0$,
the lattice is constructed recursively by replacing
each bond with $s$ parallel branches, each consisting
of $b$ bonds in series.
After $n$ iterations, the graph distance between
the two boundary vertices $A$ and $B$ is $b^n$ and the fractal dimension of the lattice is
\[
d=\frac{\log bs}{\log b}=1+\frac{\log s}{\log b}.
\]
The Hamiltonian of an elementary cell is
\begin{equation}
\mathcal H_{\mathrm{unit}}
=
-\sum_{i=1}^{s}\sum_{j=1}^{b}
J_{i j}
\sigma_{i,j-1}\sigma_{i,j},
\end{equation}
where $\sigma_{i,j}\in\{\pm1\}$ are Ising spins.
The endpoints of each branch are identified with
the boundary spins:
\[
\sigma_{i,0}=\sigma_L,
\qquad
\sigma_{i,b}=\sigma_R.
\]
The microscopic couplings are independent and
identically distributed Gaussian random variables
with zero mean and unit variance, $J_{ij}\sim\mathcal N(0,1)$.
The Hamiltonian on the generation-$n$ lattice is
\[
H_n=-\sum_{e=\{u,v\}\in E^{(n)}}
J_e\sigma_u\sigma_v ,
\]
where $E^{(n)}$ denotes the edge set of the
generation-$n$ lattice.
We impose free boundary conditions at the
two external vertices $A$ and $B$.

For a fixed realization of the couplings,
we denote the thermal expectation of an observable
$\mathcal O$ by
\[
\langle\mathcal O\rangle_n
=
\frac{1}{Z_n}
\sum_{\boldsymbol{\sigma}}
\mathcal O(\boldsymbol{\sigma})
e^{-\beta H_n(\boldsymbol{\sigma})},
\]
where $Z_n$ is the partition function and
$\beta$ is the inverse temperature.
The disorder average over the independent
Gaussian couplings is denoted by $\mathbb E$.

The hierarchical structure allows an exact
renormalization-group (RG) transformation
obtained by tracing out the internal spins
while retaining the two boundary spins.
The effective couplings obey the recursion relations~\cite{jorg2012nature}
\begin{align}
\Phi_\beta(x_1,\ldots,x_b)
&=
\frac{1}{\beta}
\operatorname{arctanh}\left(
\prod_{j=1}^{b}\tanh(\beta x_j)
\right),
\label{Phi}
\\
J_{n+1}(\beta)
&\overset{d}{=}
\sum_{i=1}^{s}
\Phi_\beta(J_{n,i,1}(\beta),\ldots,J_{n,i,b}(\beta)),
\label{Phi-sum}
\end{align}
where $J_n(\beta)$ denotes the effective coupling after
$n$ RG steps, and
$\{J_{n,i,j}(\beta)\}_{1\le i\le s,\,1\le j\le b}$
are independent copies of $J_n(\beta)$.
Here, $\overset{d}{=}$ denotes equality in distribution.
The function $\Phi_\beta$ represents the
series combination of $b$ bonds, while the
sum in Eq.~\eqref{Phi-sum} represents the
parallel combination of $s$ branches.
The initial coupling satisfies
$J_0\sim\mathcal N(0,1)$.
In the zero-temperature limit, the RG recursion
reduces to
\begin{align}
\Phi_\infty(x_1,\ldots,x_b)
&=
\operatorname{sgn}\left(
\prod_{j=1}^{b}x_j
\right)
\min_{1\le j\le b}|x_j|,
\\
G_{n+1}
&\overset{d}{=}
\sum_{i=1}^{s}
\Phi_\infty(G_{n,i,1},\ldots,G_{n,i,b}),
\label{eq:rg_zero}
\end{align}
where $G_n$ denotes the zero-temperature
effective coupling after $n$ RG steps.

We denote the finite-temperature RG transformation by
$\RG_\beta$, so that
\begin{align}
J_{n+1}(\beta) &= \RG_\beta(J_n(\beta)),\\
J_n(\beta) &= (\RG_\beta)^n(J_0).
\end{align}
Similarly, we denote the zero-temperature RG transformation
by $\RG_\infty$ and write
\begin{align}
G_{n+1} &= \RG_\infty(G_n),\\
G_n &= (\RG_\infty)^n(G_0).
\end{align}
Here, the RG transformations act on the coupling distributions,
and the equalities are understood in distribution.
Both transformations start from the same standard Gaussian
random variable $G_0=J_0=G\sim\mathcal{N}(0,1)$.

Let $\phi(x)=(2\pi)^{-1/2}e^{-x^2/2}$
denote the standard Gaussian density,
and let $g_n(x)$ denote the density of $G_n$.
We define
\begin{equation}
\alpha_n=\frac{\phi(0)}{g_n(0)}
\qquad(n\ge0).
\end{equation}
In particular, $\alpha_0=1$.
As shown below, the limit of $\alpha_n^{1/n}$
exists as $n\to\infty$ and satisfies
\begin{equation}
\Lambda(b,s)
:=
\lim_{n\to\infty}\alpha_n^{1/n}
=
\sup_{n\ge 1}\alpha_n^{1/n}.
\end{equation}
The quantity $\Lambda(b,s)$ will play a
central role in characterizing spin-glass
order on MK hierarchical lattices.

To characterize finite-temperature spin-glass
order, we introduce the site overlap between
two independent replicas:
\begin{equation}
Q_n
=
\frac{1}{|V^{(n)}|}
\sum_{i\in V^{(n)}}
\sigma_i^{(1)}\sigma_i^{(2)},
\end{equation}
where $V^{(n)}$ denotes the vertex set of the
generation-$n$ lattice, and
$\sigma^{(1)}$ and $\sigma^{(2)}$ are independent
replicas drawn from the same Gibbs measure
for a fixed disorder realization.
For observables involving two replicas,
$\langle\cdot\rangle_n$ denotes the expectation
with respect to the product Gibbs measure.
The spin-glass order parameter is then defined by
\begin{equation}
\mathbb E[\langle Q_n^2\rangle_n]
=
\frac{1}{|V^{(n)}|^2}
\sum_{i,j\in V^{(n)}}
\mathbb E\left[
\langle\sigma_i\sigma_j\rangle_n^2
\right].
\end{equation}
We characterize the presence of
finite-temperature spin-glass order by
\begin{equation}
\liminf_{n\to\infty}
\mathbb E[\langle Q_n^2\rangle_n]>0,
\end{equation}
whereas
\[
\lim_{n\to\infty}
\mathbb E[\langle Q_n^2\rangle_n]=0
\]
characterizes its absence.

At zero temperature, we instead characterize
the RG flow through the effective coupling $G_n$.
Growth of the characteristic magnitude of $G_n$
signals a flow toward the strong-coupling regime,
whereas its decay to zero indicates vanishing
stiffness at large length scales.
The zero-temperature RG flow, characterized by \(\Lambda(b,s)\), provides the basis for the criteria for the existence and absence of finite-temperature spin-glass order established below.

\begin{theorem}[Absence of finite-temperature spin-glass order]
\label{thm:sg_absence}
If $\Lambda(b,s)\le 1$, then, for every finite
inverse temperature $0<\beta<\infty$,
\begin{equation}
\lim_{n\to\infty}
\mathbb E[\langle Q_n^2\rangle_n]=0.
\end{equation}
Consequently, no finite-temperature spin-glass
order exists.

Each of the following conditions is sufficient
for $\Lambda(b,s)\le 1$:
\begin{align}
6s &< (b+1)(b+2),
\label{eq:absence_bound1}\\
\pi s &< b^2,
\label{eq:absence_bound2}\\
24s &\le 7(b+1)(b+2)
\qquad (b\ge 6).
\label{eq:absence_bound3}
\end{align}

Moreover, $\Lambda(b,s)<1$ for all integer
pairs $(b,s)$ satisfying
\begin{equation}
d=1+\frac{\log s}{\log b}
<
1+\frac{\log 17}{\log 6}
\approx 2.581.
\label{eq:sg_absence_dimension}
\end{equation}
\end{theorem}

Our second main result provides a sufficient
condition for the existence of finite-temperature
spin-glass order.

\begin{theorem}[Existence of finite-temperature spin-glass order]
\label{thm:sg_existence}
If $\Lambda(b,s)>1$, then there exists a finite
inverse temperature $\beta_{\mathrm{SG}}>0$ such that,
for every $\beta>\beta_{\mathrm{SG}}$,
\begin{equation}
\liminf_{n\to\infty}
\mathbb E[\langle Q_n^2\rangle_n]>0.
\end{equation}
Consequently, finite-temperature spin-glass
order exists at sufficiently low temperatures.

In particular, $\Lambda(b,s)>1$ whenever
\begin{equation}
\frac{b}{\sqrt{s}}
\left(
\frac{b\sqrt{s(s-1)}}{bs-1}
\right)^{s-1}<1.
\label{eq:existence_bound}
\end{equation}
Moreover, this condition holds for all
$b,s\ge2$ with fractal dimension
$d=1+\log s/\log b\ge3$.
Thus, finite-temperature spin-glass order
exists at sufficiently low temperatures
on every MK hierarchical lattice with $d\ge3$.
\end{theorem}

%


Theorem~\ref{thm:sg_absence} guarantees the
absence of finite-temperature spin-glass order
when $\pi s<b^2$.
Using $s=b^{d-1}$, this condition becomes
$\pi b^{d-3}<1$, which holds for sufficiently
large $b$ whenever $d<3$.
Together with the existence result at $d=3$,
this establishes $d=3$ as the lower critical
dimension in the large-$b$ limit of the
MK hierarchical-lattice model, consistent
with previous predictions based on scaling theory~\cite{boettcher2005stiffness}.

In the intermediate regime
\[
1+\frac{\log 17}{\log 6}\le d<3,
\]
the classification generally requires determining
whether $\Lambda(b,s)>1$ or $\Lambda(b,s)\le 1$.
The following theorem provides a finite-step
criterion for determining the sign of
$\Lambda(b,s)-1$.

\begin{theorem}[Finite-step criterion]
\label{thm:finite_step}
Let
\begin{equation}
C(b,s)=4\lceil\log_2(bs)\rceil. \label{C(b,s)}
\end{equation}
If $\alpha_n>1$ for some $n\ge 1$, then
$\Lambda(b,s)>1$.
Conversely, if $C(b,s) n\alpha_n<1$ for some
$n\ge 1$, then $\Lambda(b,s)<1$.

Furthermore, if $\Lambda(b,s)\neq 1$,
at least one of these two conditions is
satisfied for every $n$ such that
\begin{equation}
\exp\left(n|\log\Lambda(b,s)|\right)>C(b,s)n.
\end{equation}
\end{theorem}
%
%
\begin{remark}

For arbitrary integers $b,s\ge2$, one may
in principle use validated numerical
computations of the zero-temperature RG equations,
increasing $n$ until either $\alpha_n>1$
or $C(b,s) n\alpha_n<1$ is rigorously verified.
Theorem 3 guarantees that one of these
conditions is eventually satisfied whenever
$\Lambda(b,s)\neq1$.
However, the computational cost may become
substantial when $\Lambda(b,s)$ is close to
unity or when $b$ and $s$ are large.

Developing more efficient numerical methods
and sharper analytical bounds on
$\Lambda(b,s)$ remains an important
direction for future research.
\end{remark}

\begin{remark}

Whether $\Lambda(b,s)=1$ can occur
for some integers $b,s\ge2$ remains an
open question.
This marginal case corresponds to a
zero-temperature RG flow with neither
exponential growth nor exponential decay
of the characteristic coupling scale,
suggesting proximity to the lower critical dimension.
Although Theorem~\ref{thm:sg_absence}
guarantees the absence of finite-temperature
spin-glass order in this case,
neither stopping criterion can be satisfied.
Consequently, the finite-step procedure
does not terminate when $\Lambda(b,s)=1$.
\end{remark}

\section{Auxiliary Results}

In this section, we establish several auxiliary
results used in the proofs of the main theorems.

\subsection{Preservation of symmetric unimodality}
We first show that symmetric unimodality is
preserved under the RG transformations.
\begin{lemma}
\label{preservation-unimodality}
For any inverse temperature $0<\beta\le\infty$,
both the series and parallel updates preserve
symmetric unimodality, provided that the input
couplings are independent and have symmetric
unimodal densities.
Here, $\beta=\infty$ corresponds to zero temperature.
\end{lemma}

\begin{proof}
We first consider the series update with $b=2$
at finite inverse temperature $0<\beta<\infty$.
Let $f$ and $g$ denote the probability densities
of two independent input couplings $X_1$ and $X_2$,
respectively, and let $h$ denote the density
of the effective coupling
\[
X'=
\frac{1}{\beta}\operatorname{arctanh}
\left[
\tanh(\beta X_1)\tanh(\beta X_2)
\right].
\]
Since the input densities are symmetric,
so is $h$.
It therefore suffices to show that $h$ is
nonincreasing on the positive half-line.

For $x>0$, introduce the change of variables
\begin{align}
t&=-\log\tanh(\beta x),\\
x&=\psi_\beta(t)
:=
\frac{1}{\beta}\operatorname{arctanh}(e^{-t}),
\end{align}
where $\psi_\beta$ is strictly decreasing and
satisfies
\[
\psi_\beta'(t)
=
-\frac{1}{2\beta\sinh t}<0.
\]
Define
\begin{align}
T_i&=-\log\tanh(\beta|X_i|)
\qquad (i=1,2),\\
T'&=-\log\tanh(\beta|X'|).
\end{align}
The series update in Eq.~\eqref{Phi} then becomes
\[
T'=T_1+T_2.
\]

Let
\[
\widetilde f(t):=f(\psi_\beta(t)),
\qquad
\widetilde g(t):=g(\psi_\beta(t)).
\]
Since $f$ and $g$ are nonincreasing on
the positive half-line and $\psi_\beta$
is decreasing, both $\widetilde f$ and
$\widetilde g$ are nonnegative and
nondecreasing.

Denote the probability densities of
$T_1$, $T_2$, and $T'$ by
$p_1$, $p_2$, and $p'$, respectively.
The change-of-variables formula gives
\[
p_1(t)=\frac{\widetilde f(t)}{\beta\sinh t},
\qquad
p_2(t)=\frac{\widetilde g(t)}{\beta\sinh t},
\qquad
p'(t)=\frac{h(\psi_\beta(t))}{\beta\sinh t}.
\]
Since $T_1$ and $T_2$ are independent,
\[
p'(t)=\int_0^t p_1(v)p_2(t-v)\,dv.
\]
Consequently,
\begin{align}
h(\psi_\beta(t))
&=
\frac{1}{\beta}\int_0^t
\widetilde f(v)\widetilde g(t-v)
\frac{\sinh t}
{\sinh v\,\sinh(t-v)}\,dv
\nonumber\\
&=
\frac{1}{\beta}
\left[
(\widetilde f\cdot\coth)*\widetilde g
+
\widetilde f*(\widetilde g\cdot\coth)
\right](t),
\label{eq:unimodal_convolution}
\end{align}
where $*$ denotes convolution on the
positive half-line and $\cdot$ denotes
pointwise multiplication.
In the second equality, we used
\[
\frac{\sinh t}{\sinh v\,\sinh(t-v)}
=
\coth v+\coth(t-v).
\]

Although $\coth t$ diverges as $t\to0$,
the functions $\widetilde f\cdot\coth$ and
$\widetilde g\cdot\coth$ are locally integrable.
Indeed,
\[
\widetilde f(t)\coth t
=
\beta p_1(t)\cosh t,
\qquad
\widetilde g(t)\coth t
=
\beta p_2(t)\cosh t.
\]
Since $p_1$ and $p_2$ are probability
densities and $\cosh t$ is bounded
on every compact interval, both
functions are nonnegative and locally
integrable on $[0,\infty)$.

We now use the fact that the convolution
of a nonnegative locally integrable
function with a nonnegative nondecreasing
function is nondecreasing.
Since $\widetilde f$ and $\widetilde g$
are nonnegative and nondecreasing,
Eq.~\eqref{eq:unimodal_convolution}
implies that $h(\psi_\beta(t))$ is
nondecreasing in $t$.
As $\psi_\beta$ is strictly decreasing,
$h(x)$ is nonincreasing for $x>0$.
Thus, symmetric unimodality is preserved
under the series update with $b=2$.
The result extends to arbitrary $b$
by the associativity of the series update.

At zero temperature ($\beta=\infty$),
the series update with $b=2$ reduces to
\[
X'=
\operatorname{sgn}(X_1X_2)
\min\{|X_1|,|X_2|\}.
\]
For $x>0$, the output density is
\[
h(x)
=
f(x)\mathbb P(|X_2|\ge x)
+
g(x)\mathbb P(|X_1|\ge x).
\]
Since all factors on the right-hand side
are nonnegative and nonincreasing in $x$,
$h$ is nonincreasing on the positive
half-line.
The result again extends to arbitrary $b$
by associativity.

Finally, the parallel update in
Eq.~\eqref{Phi-sum} corresponds to the
sum of independent couplings.
The convolution of two symmetric unimodal
densities is again symmetric unimodal~\cite{feller1971introduction}.
Repeated convolution therefore establishes
the result for any $s$.
\end{proof}

Since the initial couplings
$J_0\overset{d}{=}G_0\sim\mathcal N(0,1)$
have a symmetric unimodal density,
Lemma~\ref{preservation-unimodality}
implies that the distributions of
$J_n$ and $G_n$ remain symmetric unimodal
at every generation $n$.

\subsection{Preservation of peakedness order}

Next, we introduce the peakedness order for
symmetric random variables, defined by
\begin{equation}
X\preceq Y
\quad\Longleftrightarrow\quad
\mathbb P(|X|\le t)
\ge
\mathbb P(|Y|\le t)
\quad (t\ge 0).
\label{eq:order}
\end{equation}
Equivalently, $|X|$ is smaller than $|Y|$
in the usual stochastic order.
Thus, $X\preceq Y$ means that $X$ is more
concentrated around the origin than $Y$.

Using Lemma~\ref{preservation-unimodality},
we establish that the RG transformations
preserve the peakedness order.

\begin{lemma}
\label{lem:order}
Let $X$ and $Y$ be random variables with
symmetric unimodal densities satisfying
$X\preceq Y$.
Then, for every $0<\beta\le\infty$,
\begin{equation}
\RG_\beta(X)\preceq\RG_\beta(Y).
\end{equation}
Moreover, for every $0<\beta<\infty$,
\begin{equation}
\RG_\beta(X)\preceq\RG_\infty(X).
\label{eq:order-rg}
\end{equation}
\end{lemma}
\begin{proof}
We first consider the series update.
For every $0<\beta\le\infty$,
\[
|\Phi_\beta(x_1,\ldots,x_b)|
=
\Phi_\beta(|x_1|,\ldots,|x_b|),
\]
where $\Phi_\beta$ is nondecreasing in
each argument on $[0,\infty)^b$.
For $\beta=\infty$, this follows directly
from the definition of $\Phi_\infty$.

Let $F_X$ and $F_Y$ denote the distribution
functions of $|X|$ and $|Y|$, respectively.
Since $X\preceq Y$, we have $F_X\ge F_Y$.
For independent uniform random variables
$U_1,\ldots,U_b$ on $(0,1)$, define
\[
A_i=F_X^{-1}(U_i),
\qquad
B_i=F_Y^{-1}(U_i),
\]
where $F^{-1}$ denotes the generalized inverse
distribution function.
Then $A_i\le B_i$ almost surely, and hence
\[
\Phi_\beta(A_1,\ldots,A_b)
\le
\Phi_\beta(B_1,\ldots,B_b)
\]
almost surely.
Since the $A_i$ and $B_i$ are independent
copies of $|X|$ and $|Y|$, respectively,
we obtain
\[
\Phi_\beta(X_1,\ldots,X_b)
\preceq
\Phi_\beta(Y_1,\ldots,Y_b),
\]
where the $X_i$ and $Y_i$ are independent
copies of $X$ and $Y$, respectively.
Furthermore, for every $0<\beta<\infty$,
\[
|\Phi_\beta(x_1,\ldots,x_b)|
\le
\min_{1\le i\le b}|x_i|
=
|\Phi_\infty(x_1,\ldots,x_b)|.
\]
Therefore,
\[
\Phi_\beta(X_1,\ldots,X_b)
\preceq
\Phi_\infty(X_1,\ldots,X_b).
\]

The peakedness order is preserved under
convolution of independent symmetric unimodal
random variables, as established by
Birnbaum~\cite{birnbaum1948random}.
By Lemma~\ref{preservation-unimodality},
the series outputs are symmetric unimodal.
Applying this convolution property to the
$s$ independent branches in
Eq.~\eqref{Phi-sum}, we obtain
\[
\RG_\beta(X)\preceq\RG_\beta(Y)
\qquad(0<\beta\le\infty),
\]
and
\[
\RG_\beta(X)\preceq\RG_\infty(X)
\qquad(0<\beta<\infty).
\]
This completes the proof.

%
\end{proof}

Since both RG flows start from the same standard
Gaussian distribution, Lemma~\ref{lem:order}
implies, by induction, that
\begin{equation}
J_n\preceq G_n
\qquad(n\ge0)
\label{eq:finite_zero_order}
\end{equation}
for every finite inverse temperature
$0<\beta<\infty$.

\subsection{Preservation of Gaussian scale mixtures at zero temperature}

We next introduce the class of Gaussian scale mixtures, which is preserved under the zero-temperature
RG transformation.
A random variable $X$ is called a Gaussian scale mixture
if it admits the representation
\[
X=\sqrt{V}\,G,
\]
where $G\sim\mathcal{N}(0,1)$ and $V>0$ is a
random variable independent of $G$.

\begin{lemma}
\label{lem:gsm}
Let $X=\sqrt{V}\,G$, where
$G\sim\mathcal N(0,1)$ and $V>0$ is a
random variable independent of $G$.
Then $\RG_\infty(X)$ is also a centered
Gaussian scale mixture with a strictly
positive mixing variance.
\end{lemma}

\begin{proof}
Let $p$ denote the probability density of
$X$, and define
\[
S(t):=\mathbb P(|X|>t)
\qquad(t\ge0).
\]
For $z>0$, we have
\begin{align}
p(\sqrt{2z})
&=
\frac{1}{\sqrt{2\pi}}
\mathbb E_V\left[
V^{-1/2}e^{-z/V}
\right],
\\
S(\sqrt{2z})
&=
\mathbb E_V\left[
\operatorname{erfc}\sqrt{z/V}
\right],
\end{align}
where $\mathbb E_V$ denotes expectation
with respect to the mixing variable $V$.
Using the identity
\begin{equation}
\operatorname{erfc}\sqrt{z}
=
\int_1^\infty
e^{-uz}\frac{du}{\pi u\sqrt{u-1}},
\qquad z>0,
\label{eq:erfc}
\end{equation}
we find that both $p(\sqrt{2z})$ and
$S(\sqrt{2z})$ are Laplace transforms
of positive measures supported on
$(0,\infty)$.

We first consider the zero-temperature
series update of $b$ independent copies
$X_1,\ldots,X_b$ of $X$.
By the definition of $\Phi_\infty$,
\[
|\Phi_\infty(X_1,\ldots,X_b)|
=
\min_{1\le i\le b}|X_i|.
\]
Since the input couplings are symmetric
and have no atom at zero, their signs
are independent of their magnitudes
and are equally likely to be positive
or negative.
Thus, the output has a symmetric density
$q$ given by
\[
q(x)=b\,p(x)S(x)^{b-1},
\qquad x>0.
\]
Since products of Laplace transforms
of positive measures are again Laplace
transforms of positive measures, there
exists a positive measure $\nu$ supported
on $(0,\infty)$ such that
\[
q(x)=
\int_0^\infty e^{-ux^2/2}\,\nu(du).
\]
By symmetry, this representation holds
for almost every $x\in\mathbb R$.
Since $q$ is a probability density,
Tonelli's theorem gives
\[
\begin{aligned}
1
&=\int_{-\infty}^{\infty}q(x)\,dx\\
&=\int_0^\infty
\left(
\int_{-\infty}^{\infty}
e^{-ux^2/2}\,dx
\right)\nu(du)\\
&=\int_0^\infty
\sqrt{\frac{2\pi}{u}}\,\nu(du).
\end{aligned}
\]
Hence,
\[
\mu(du):=
\sqrt{\frac{2\pi}{u}}\,\nu(du)
\]
defines a probability measure on
$(0,\infty)$, and
\[
q(x)=
\int_0^\infty
\sqrt{\frac{u}{2\pi}}
e^{-ux^2/2}\,\mu(du).
\]
This is a mixture of centered Gaussian
densities with variance $1/u$.
Since $\mu$ is supported on $(0,\infty)$,
the mixing variance is strictly positive
almost surely.
Thus, the series update preserves the
class of centered Gaussian scale mixtures
with strictly positive mixing variances.

Finally, the class is also preserved
under the parallel update.
Indeed, if $X_1,\ldots,X_s$ are independent
centered Gaussian scale mixtures with
strictly positive mixing variances
$V_1,\ldots,V_s$, then
\[
\sum_{i=1}^s X_i
\overset{d}{=}
\sqrt{\sum_{i=1}^s V_i}\,G',
\]
where $G'\sim\mathcal N(0,1)$ is
independent of the mixing variables.
Since $\sum_{i=1}^s V_i>0$ almost surely,
the parallel update also preserves
the class.

Combining the series and parallel updates
proves that the class is preserved under
the zero-temperature MK RG transformation
$\RG_\infty$.
\end{proof}

Since the initial coupling
$G_0\sim\mathcal N(0,1)$ is a centered
Gaussian scale mixture with unit variance,
Lemma~\ref{lem:gsm} implies that the
zero-temperature effective coupling $G_n$
remains a centered Gaussian scale mixture
with a strictly positive mixing variance
at every generation $n$.

\subsection{Supermultiplicativity and asymptotic growth of $\alpha_n$}

Recall that $G_n=(\RG_\infty)^n(G)$ denotes
the zero-temperature effective coupling,
with probability density $g_n(x)$, and that
\[
\alpha_n=\frac{1}{\sqrt{2\pi}\,g_n(0)}.
\]
Lemmas~\ref{lem:order} and~\ref{lem:gsm}
lead to the following key property.

\begin{lemma}
\label{lem:supermultiplicativity}
For all integers $n,m\ge0$,
\begin{equation}
\alpha_{n+m}\ge\alpha_n\alpha_m.
\end{equation}
\end{lemma}
\begin{proof}
By Lemma~\ref{lem:gsm}, the class of centered
Gaussian scale mixtures with strictly positive
mixing variances is preserved under $\RG_\infty$.
Since $G_0\sim\mathcal N(0,1)$, each $G_n$
admits the representation
\[
G_n\overset{d}{=}\sqrt{V_n}\,Z,
\]
where $Z\sim\mathcal N(0,1)$ and $V_n>0$
is independent of $Z$.

We first note that $g_n(0)$ is finite and
strictly positive for every $n$.
Indeed, the zero-temperature series update
maps a symmetric unimodal density $p$
to a density $q$ satisfying
\[
q(x)=b\,p(x)S(|x|)^{b-1}
\qquad (x\ne0),
\]
where $S(t)=\mathbb P(|X|>t)$.
Thus, $\|q\|_\infty\le b\|p\|_\infty$.
The subsequent parallel update is a
convolution of probability densities,
which cannot increase the supremum norm.
Since $g_0(0)=1/\sqrt{2\pi}$, induction gives
\[
0<g_n(0)\le\frac{b^n}{\sqrt{2\pi}}.
\]
The densities are also continuous at the origin.

Writing $\phi(0)=1/\sqrt{2\pi}$,
the Gaussian scale-mixture representation yields
\[
g_n(0)
=
\phi(0)\mathbb E_{V_n}[V_n^{-1/2}],
\]
where $\mathbb E_{V_n}$ denotes expectation
with respect to the mixing variable $V_n$.
Consequently,
\[
\mathbb E_{V_n}[V_n^{-1/2}]
=
\frac{1}{\alpha_n}.
\]
Let $F_Z(u)=\mathbb P(|Z|\le u)$ denote
the distribution function of $|Z|$.
Since $F_Z$ is concave on $[0,\infty)$,
Jensen's inequality gives, for $t\ge0$,
\begin{align}
\mathbb P(|G_n|\le t)
&=
\mathbb E_{V_n}
\left[F_Z(tV_n^{-1/2})\right]
\nonumber\\
&\le
F_Z\left(
t\mathbb E_{V_n}[V_n^{-1/2}]
\right)
\nonumber\\
&=
F_Z(t/\alpha_n)
\nonumber\\
&=
\mathbb P(|\alpha_n Z|\le t).
\end{align}
Therefore,
\[
\alpha_n Z\preceq G_n.
\]
By Lemma~\ref{lem:order} and the positive
homogeneity of $\RG_\infty$, we obtain
\[
\begin{aligned}
G_{n+m}
&\overset{d}{=}(\RG_\infty)^m(G_n)\\
&\succeq(\RG_\infty)^m(\alpha_n Z)\\
&\overset{d}{=}\alpha_n G_m.
\end{aligned}
\]
Consequently,
\[
F_{G_{n+m}}(t)
\le F_{G_m}(t/\alpha_n),
\qquad t\ge0,
\]
where $F_X(t)=\mathbb P(|X|\le t)$.
Since the densities $g_n$ are continuous
at the origin,
\[
\lim_{t\downarrow0}
\frac{F_{G_n}(t)}{2t}
=
g_n(0).
\]
Dividing the preceding inequality by $2t$
and taking $t\downarrow0$, we obtain
\[
g_{n+m}(0)
\le
\frac{g_m(0)}{\alpha_n}.
\]
Therefore,
\[
\begin{aligned}
\alpha_{n+m}
&=
\frac{\phi(0)}{g_{n+m}(0)}\\
&\ge
\alpha_n\frac{\phi(0)}{g_m(0)}\\
&=
\alpha_n\alpha_m,
\end{aligned}
\]
which completes the proof.
\end{proof}

By Lemma~\ref{lem:supermultiplicativity},
the sequence $\{\alpha_n\}$ is
supermultiplicative.
Fekete's lemma then yields the
following result.

\begin{lemma}
\label{Asymptotic growth rate}
The limit defining $\Lambda(b,s)$ exists
and satisfies
\begin{equation}
\Lambda(b,s)
=
\lim_{n\to\infty}\alpha_n^{1/n}
=
\sup_{n\ge1}\alpha_n^{1/n}.
\end{equation}
Consequently,
\begin{equation}
\alpha_n\le\Lambda(b,s)^n
\end{equation}
for every $n\ge1$.
In particular, if $\Lambda(b,s)\le1$,
then $\alpha_n\le1$ for all $n\ge1$.
Conversely, if $\alpha_n>1$ for some
$n\ge1$, then $\Lambda(b,s)>1$.
\end{lemma}
\begin{proof}
By Lemma~\ref{lem:supermultiplicativity},
\[
\alpha_{n+m}\ge\alpha_n\alpha_m.
\]
Since $0<\alpha_n<\infty$, the sequence
$\{\log\alpha_n\}$ is superadditive.
Fekete's lemma therefore gives
\[
\lim_{n\to\infty}
\frac{\log\alpha_n}{n}
=
\sup_{n\ge1}
\frac{\log\alpha_n}{n}.
\]

To verify that the limit is finite,
we use the Gaussian scale-mixture
representation of $G_n$.
By Jensen's inequality,
\[
\alpha_n
=
\frac{1}{\mathbb E_{V_n}[V_n^{-1/2}]}
\le
\sqrt{\mathbb E_{V_n}[V_n]}
=
\sqrt{\mathbb E[G_n^2]}.
\]
Since the zero-temperature series update
satisfies
\[
|\Phi_\infty(x_1,\ldots,x_b)|
=
\min_{1\le j\le b}|x_j|,
\]
and the independent branch couplings
are symmetric, we have
\[
\mathbb E[G_{n+1}^2]
\le s\,\mathbb E[G_n^2].
\]
Together with $\mathbb E[G_0^2]=1$,
this yields
\[
\alpha_n\le s^{n/2}.
\]
Thus, the limit obtained from Fekete's
lemma is finite.

Exponentiating gives
\[
\Lambda(b,s)
=
\lim_{n\to\infty}\alpha_n^{1/n}
=
\sup_{n\ge1}\alpha_n^{1/n}.
\]
The remaining assertions follow
immediately from the supremum
representation.
\end{proof}

\section{Proof of Theorem~\ref{thm:sg_absence}}

We first show that if $\Lambda(b,s)<1$,
the zero-temperature effective coupling
vanishes under the RG flow, indicating
the absence of large-scale stiffness.
We then use the comparison between
finite- and zero-temperature effective
couplings to analyze finite-temperature
spin-glass order.

The key observation is that symmetric
log-concavity is preserved under the
zero-temperature RG transformation.

\subsection{Preservation of symmetric log-concavity at zero temperature}

\begin{lemma}
\label{lem:logconcavity}
The zero-temperature MK RG transformation
$\RG_\infty$ preserves the class of symmetric
log-concave probability densities.
\end{lemma}
\begin{proof}
Let $p$ denote the symmetric log-concave
density of the input coupling $X$, and define
\[
S(t):=\mathbb P(|X|>t)
=2\int_t^\infty p(x)\,dx,
\qquad t\ge0.
\]
By Pr\'ekopa's theorem~\cite{prekopa1973logarithmic},
the log-concavity of $p$ implies that
$S$ is log-concave on $[0,\infty)$.

We first consider the zero-temperature
series update of $b$ independent copies
of $X$.
The output has a symmetric density
\[
q(x)=b\,p(x)S(|x|)^{b-1}.
\]
Since $S$ is nonincreasing and log-concave
on $[0,\infty)$, the function
$x\mapsto S(|x|)$ is log-concave on
$\mathbb R$.
Consequently, $q$ is symmetric and
log-concave, being the product of
nonnegative log-concave functions.

The parallel update corresponds to the
sum of $s$ independent couplings.
Since convolution preserves log-concavity
by Pr\'ekopa's theorem~\cite{vandenberghe2004convex}, and also preserves
symmetry, the resulting density is again
symmetric and log-concave.

Thus, both the series and parallel updates
preserve symmetric log-concavity, proving
the lemma.
\end{proof}

Unlike the zero-temperature transformation,
the finite-temperature RG transformation
does not generally preserve log-concavity.

\subsection{Uniform bounds on finite-temperature couplings}

The zero-temperature preservation property
will allow us to derive an exponential
tail bound for the effective coupling $G_n$.

\begin{lemma}
\label{lem:tail}
Let $c_0=\sqrt{2/\pi}$. Then, for every
$n\ge0$ and $t\ge0$,
\begin{equation}
\mathbb P(|G_n|>t)
\le
\exp\left(
-\frac{c_0t}{\Lambda(b,s)^n}
\right).
\end{equation}
\end{lemma}
\begin{proof}
Define the survival function
\[
S_n(t):=\mathbb P(|G_n|>t)
=2\int_t^\infty g_n(x)\,dx,
\qquad t\ge0.
\]
By Lemma~\ref{lem:logconcavity},
$g_n$ is symmetric and log-concave.
Pr\'ekopa's theorem~\cite{prekopa1973logarithmic}, therefore implies
that $S_n$ is also log-concave.

Since $S_n(0)=1$ and $g_n$ is continuous
at the origin, we have
\[
\log S_n(0)=0,
\qquad
(\log S_n)'(0+)=-2g_n(0).
\]
By the concavity of $\log S_n$,
\[
\log S_n(t)\le-2g_n(0)t.
\]
Hence,
\begin{align}
\mathbb P(|G_n|>t)
&\le \exp\left(-2g_n(0)t\right)
\nonumber\\
&=\exp\left(
-\frac{c_0t}{\alpha_n}
\right),
\end{align}
where $c_0=2/\sqrt{2\pi}=\sqrt{2/\pi}$.

By Lemma~\ref{Asymptotic growth rate},
$\alpha_n\le\Lambda(b,s)^n$.
Therefore,
\[
\mathbb P(|G_n|>t)
\le
\exp\left(
-\frac{c_0t}{\Lambda(b,s)^n}
\right),
\]
which completes the proof.
\end{proof}

Consequently, if $\Lambda(b,s)<1$,
the zero-temperature effective coupling
$G_n$ converges to zero in probability
as $n\to\infty$.
In particular, for every fixed $t>0$,
the probability $\mathbb P(|G_n|>t)$
decays at least super-exponentially
with $n$.

We next use the peakedness comparison
between $J_n$ and $G_n$ to extend this
bound to the finite-temperature RG flow.

\begin{lemma}
\label{lem:finite_tail}
For every finite inverse temperature
$0<\beta<\infty$ and every $n\ge0$,
the effective coupling $J_n(\beta)$ satisfies
\begin{align}
\mathbb P(|J_n(\beta)|>t)
&\le
\exp\left(
-\frac{c_0t}{\Lambda(b,s)^n}
\right),
\qquad t\ge0,
\label{eq:gaussian-upper}\\
\mathbb E[J_n(\beta)^2]
&\le
\pi\Lambda(b,s)^{2n},
\label{eq:gaussian-moment}
\end{align}
where $J_n(\beta)$ denotes the previously defined
finite-temperature effective coupling $J_n$,
with its dependence on $\beta$ made explicit.
\end{lemma}
\begin{proof}
Since $J_0(\beta)\overset{d}{=}G_0$,
we have $J_0(\beta)\preceq G_0$.
Suppose that $J_n(\beta)\preceq G_n$.
By Lemma~\ref{lem:order},
\begin{align}
J_{n+1}(\beta)
&\overset{d}{=}
\RG_\beta(J_n(\beta))
\nonumber\\
&\preceq
\RG_\infty(J_n(\beta))
\nonumber\\
&\preceq
\RG_\infty(G_n)
\overset{d}{=}G_{n+1}.
\end{align}
Thus, by induction,
\begin{equation}
J_n(\beta)\preceq G_n
\qquad(n\ge0,\ 0<\beta<\infty).
\label{eq:directcompare}
\end{equation}
Combining Eq.~\eqref{eq:directcompare}
with the exponential tail bound in
Lemma~\ref{lem:tail}, we obtain
\[
\begin{aligned}
\mathbb P(|J_n(\beta)|>t)
&\le\mathbb P(|G_n|>t)\\
&\le
\exp\left(
-\frac{c_0t}{\Lambda(b,s)^n}
\right).
\end{aligned}
\]
Furthermore, the peakedness order implies
\[
\begin{aligned}
\mathbb E[J_n(\beta)^2]
&\le\mathbb E[G_n^2]\\
&=2\int_0^\infty
t\,\mathbb P(|G_n|>t)\,dt\\
&\le
2\int_0^\infty
t\exp\left(
-\frac{c_0t}{\Lambda(b,s)^n}
\right)dt\\
&=
\frac{2}{c_0^2}\Lambda(b,s)^{2n}\\
&=\pi\Lambda(b,s)^{2n},
\end{aligned}
\]
where we used $c_0=\sqrt{2/\pi}$.
This proves both bounds.
\end{proof}

In particular, if $\Lambda(b,s)<1$,
Eq.~\eqref{eq:gaussian-moment} implies
\[
\lim_{n\to\infty}
\mathbb E[J_n(\beta)^2]=0
\]
uniformly in $0<\beta<\infty$.
Thus, the finite-temperature effective
coupling vanishes in mean square.

For $\Lambda(b,s)\le1$,
Eq.~\eqref{eq:gaussian-upper} yields
the uniform tail bound
\begin{equation}
\sup_{\substack{n\ge0\\0<\beta<\infty}}
\mathbb P(|J_n(\beta)|>M)
\le e^{-c_0M}
\xrightarrow[M\to\infty]{}0.
\label{eq:tight}
\end{equation}
This establishes uniform tightness
of the effective-coupling distributions
over all generations and finite
inverse temperatures.
The same bounds hold for the
zero-temperature effective coupling $G_n$.

\subsection{Absence of finite-temperature spin-glass phase}

These estimates provide control over
the finite-temperature RG flow.
To establish the absence of spin-glass
order, we must further relate the
effective couplings to correlations
between spins on the hierarchical lattice.
We next examine the influence of fixed
boundary spins on the magnetization
of interior vertices.

\begin{lemma}
\label{lem:boundary_decay}
Fix an inverse temperature $0<\beta<\infty$.
Let $A$ and $B$ denote the two boundary
vertices of the generation-$k$ MK lattice,
and define
\[
V_{\mathrm{int}}^{(k)}
:=V^{(k)}\setminus\{A,B\},
\qquad
I_k:=|V_{\mathrm{int}}^{(k)}|.
\]
For $k\ge1$, let
\[
B_k:=
\frac{1}{I_k}
\sum_{v\in V_{\mathrm{int}}^{(k)}}
\mathbb E[h_k(v;\boldsymbol J)],
\]
where
\[
h_k(v;\boldsymbol J):=
\max_{\eta_A,\eta_B\in\{\pm1\}}
\left|
\langle\sigma_v\rangle_k^{\eta_A,\eta_B}
\right|,
\]
and $\langle\cdot\rangle_k^{\eta_A,\eta_B}$
denotes the thermal expectation with
the boundary spins fixed to
$\sigma_A=\eta_A$ and $\sigma_B=\eta_B$.
Here, $\boldsymbol J$ denotes the collection of microscopic couplings on the lattice.
We set $B_0=0$ by convention.

If $\Lambda(b,s)\le1$, then
\begin{equation}
\lim_{k\to\infty}B_k=0.
\end{equation}
\end{lemma}
\begin{proof}
Consider a generation-$k$ child cell $C$
with endpoint spins $\sigma_x$ and $\sigma_y$.
For an interior vertex $v$ of $C$, define
the conditional magnetization
\[
m_{C,v}(\sigma_x,\sigma_y)
:=
\langle\sigma_v\mid\sigma_x,\sigma_y\rangle_C.
\]
Since there is no external magnetic field,
global spin-flip symmetry implies
\[
m_{C,v}(\sigma_x,\sigma_y)
=
-m_{C,v}(-\sigma_x,-\sigma_y).
\]
Consequently, there exist coefficients
$\ell_{C,v}$ and $r_{C,v}$ such that
\[
m_{C,v}(\sigma_x,\sigma_y)
=
\ell_{C,v}\sigma_x+r_{C,v}\sigma_y.
\]
These coefficients depend only on the
microscopic couplings within $C$, and
\begin{equation}
h_k(v;\boldsymbol J_C)
=
|\ell_{C,v}|+|r_{C,v}|
\le1,
\label{eq:boundary_response}
\end{equation}
where $\boldsymbol J_C$ denotes the collection of couplings within the cell $C$.

We now consider a generation-$(k+1)$
parent cell, consisting of $s$ parallel
branches with $b$ generation-$k$ child
cells in series along each branch.
The parent cell contains $bs$ child cells
and $s(b-1)$ additional internal vertices,
which we call skeleton vertices.

Denote the child cells by $C_{i,j}$,
where $1\le i\le s$ and $1\le j\le b$.
Tracing out the internal spins of each
child cell yields an effective coupling
$K_{i,j}^{(k)}$ between its endpoints.
These couplings are independent copies
of $J_k(\beta)$:
\[
K_{i,j}^{(k)}
\overset{d}{=}J_k(\beta).
\]
The resulting skeleton is an Ising model
with these effective couplings.

For $M>0$, define
\[
\rho_M:=\tanh(2\beta M)<1,
\qquad
\varepsilon_M:=
bs\sup_{j\ge0}
\mathbb P(|J_j(\beta)|>M).
\]
Let $\mathcal A_{k,M}$ be the event that
all $bs$ effective couplings satisfy
\[
|K_{i,j}^{(k)}|\le M.
\]
By the union bound,
\begin{equation}
\mathbb P(\mathcal A_{k,M}^c)
\le
bs\,\mathbb P(|J_k(\beta)|>M)
\le\varepsilon_M.
\label{eq:boundary_bad_event}
\end{equation}

On $\mathcal A_{k,M}$, consider an internal
skeleton vertex $w$ with neighboring
skeleton spins $\sigma_1$ and $\sigma_2$,
connected through effective couplings
$K_1$ and $K_2$, respectively.
Its conditional magnetization satisfies
\[
\begin{aligned}
|\langle\sigma_w\mid\sigma_1,\sigma_2
\rangle_{\mathrm{skel}}|
&=
|\tanh[\beta(K_1\sigma_1+K_2\sigma_2)]|\\
&\le\tanh(2\beta M)
=\rho_M.
\end{aligned}
\]
Averaging over the neighboring spins gives
\begin{equation}
|\langle\sigma_w\rangle_{\mathrm{skel}}|
\le\rho_M.
\label{eq:skeleton_magnetization}
\end{equation}
This bound holds uniformly over the
fixed boundary conditions of the parent cell.

Next, consider an interior vertex $v$
of a child cell $C$ with endpoints $x$ and $y$.
The conditional magnetization representation
gives
\[
\langle\sigma_v\rangle_{k+1}
=
\ell_{C,v}\langle\sigma_x\rangle_{\mathrm{skel}}
+
r_{C,v}\langle\sigma_y\rangle_{\mathrm{skel}},
\]
where the fixed boundary conditions of
the parent cell are understood.
Let
\[
\delta_x:=\mathbf 1_{\{x\in\{A,B\}\}},
\qquad
\delta_y:=\mathbf 1_{\{y\in\{A,B\}\}}.
\]
On $\mathcal A_{k,M}$,
Eq.~\eqref{eq:skeleton_magnetization}
implies
\[
|\langle\sigma_x\rangle_{\mathrm{skel}}|
\le\rho_M+(1-\rho_M)\delta_x,
\]
and similarly for $\sigma_y$.
Consequently,
\[
\begin{aligned}
|\langle\sigma_v\rangle_{k+1}|
\le{}&
\rho_M h_k(v;\boldsymbol J_C)\\
&+(1-\rho_M)
\left(
\delta_x|\ell_{C,v}|
+\delta_y|r_{C,v}|
\right).
\end{aligned}
\]
On the complementary event
$\mathcal A_{k,M}^c$, we use
\[
|\langle\sigma_v\rangle_{k+1}|
\le h_k(v;\boldsymbol J_C)\le1.
\]
Thus, uniformly over the boundary
conditions of the parent cell,
\begin{align}
h_{k+1}(v;\boldsymbol J)
\le{}&
\rho_M h_k(v;\boldsymbol J_C)
\nonumber\\
&+(1-\rho_M)
\left(
\delta_x|\ell_{C,v}|
+\delta_y|r_{C,v}|
\right)
\nonumber\\
&+(1-\rho_M)
h_k(v;\boldsymbol J_C)
\mathbf 1_{\mathcal A_{k,M}^c}.
\label{eq:boundary_child_bound}
\end{align}

We now sum over the interior vertices
of all $bs$ child cells and take
the disorder average.
The first term contributes
\[
\rho_MbsI_kB_k.
\]

By reflection symmetry of the child-cell
geometry and the disorder distribution,
\[
\begin{aligned}
\sum_{v\in V_{\mathrm{int}}^{(k)}}
\mathbb E|\ell_{C,v}|
&=
\sum_{v\in V_{\mathrm{int}}^{(k)}}
\mathbb E|r_{C,v}|=\frac{I_kB_k}{2}.
\end{aligned}
\]
Each branch contains one child cell
adjacent to the left boundary and one
adjacent to the right boundary.
Hence, the second term contributes
at most
\[
(1-\rho_M)sI_kB_k.
\]

For the last term, using
$h_k(v;\boldsymbol J_C)\le1$ and
Eq.~\eqref{eq:boundary_bad_event}, we have
\[
\begin{aligned}
&\sum_{C}\sum_{v\in V_{\mathrm{int}}^{(k)}(C)}
\mathbb E\left[
h_k(v;\boldsymbol J_C)
\mathbf 1_{\mathcal A_{k,M}^c}
\right]\\
&\qquad\le
bsI_k\,\mathbb P(\mathcal A_{k,M}^c)\\
&\qquad\le bsI_k\varepsilon_M.
\end{aligned}
\]
No independence between
$\mathcal A_{k,M}$ and the internal
magnetizations is required.

The $s(b-1)$ skeleton vertices contribute
at most $s(b-1)$.
Combining these estimates yields
\begin{align}
I_{k+1}B_{k+1}
\le{}&
[\rho_Mbs+(1-\rho_M)s]I_kB_k
\nonumber\\
&+(1-\rho_M)bsI_k\varepsilon_M
+s(b-1).
\label{eq:boundary_recursion_raw}
\end{align}

Since
\[
I_{k+1}=bsI_k+s(b-1),
\]
we obtain
\begin{align}
B_{k+1}
\le{}&
\left(
\rho_M+\frac{1-\rho_M}{b}
\right)B_k
\nonumber\\
&+(1-\rho_M)\varepsilon_M
+\frac{s(b-1)}{I_{k+1}}.
\label{eq:boundary_recursion}
\end{align}

Since $b\ge2$ and $\rho_M<1$,
\[
\rho_M+\frac{1-\rho_M}{b}<1.
\]
Moreover, $I_k\to\infty$ as $k\to\infty$.
Taking the limit superior in
Eq.~\eqref{eq:boundary_recursion}
gives
\[
\limsup_{k\to\infty}B_k
\le
\frac{\varepsilon_M}{1-1/b}.
\]

Under the assumption $\Lambda(b,s)\le1$,
Lemma~\ref{lem:finite_tail} yields
\[
\varepsilon_M
\le bs\,e^{-c_0M}
\xrightarrow[M\to\infty]{}0.
\]
Since $M>0$ is arbitrary and $B_k\ge0$,
we conclude that
\[
\lim_{k\to\infty}B_k=0.
\]
\end{proof}

The decay of the boundary-induced
magnetization established above provides
a key ingredient for controlling
spin correlations on the MK hierarchical
lattice. We now use this result to
prove the absence of finite-temperature
spin-glass order.

\begin{lemma}
\label{lem:site_overlap}
If $\Lambda(b,s)\le1$, then, for every
finite inverse temperature $0<\beta<\infty$,
\begin{equation}
\lim_{N\to\infty}
\mathbb E[\langle Q_N^2\rangle_N]=0.
\end{equation}
\end{lemma}
\begin{proof}
Fix $k\ge1$ and consider the generation-$N$
MK lattice with $N>k$.
The generation-$N$ lattice can be obtained
by replacing each bond of the
generation-$(N-k)$ lattice with a
generation-$k$ cell.
Thus, it consists of
\[
m=(bs)^{N-k}
\]
generation-$k$ cells connected through
a skeleton corresponding to the
generation-$(N-k)$ lattice.

Let $\mathcal C_{N,k}$ denote the collection
of these cells.
For each $C\in\mathcal C_{N,k}$, let
$V_{\mathrm{int}}(C)$ denote its interior
vertex set.
Each cell contains $I_k$ interior vertices,
and these sets are mutually disjoint.
The remaining vertices form the skeleton,
whose vertex set is identified with
$V^{(N-k)}$.
Consequently,
\begin{equation}
|V^{(N)}|
=
mI_k+|V^{(N-k)}|.
\label{eq:vertex_decomposition}
\end{equation}
Consider two interior vertices
$v\in V_{\mathrm{int}}(C)$ and
$w\in V_{\mathrm{int}}(D)$ belonging to
distinct cells $C\ne D$.
Conditioned on all skeleton spins,
the interior spins of distinct cells
are independent.
Therefore,
\[
\langle\sigma_v\sigma_w\rangle_N
=
\left\langle
m_{C,v}(\sigma_{x_C},\sigma_{y_C})
m_{D,w}(\sigma_{x_D},\sigma_{y_D})
\right\rangle_{\mathrm{skel}},
\]
where $x_C,y_C$ and $x_D,y_D$ denote the
endpoints of cells $C$ and $D$, respectively,
$m_{C,v}$ and $m_{D,w}$ are the conditional
magnetizations defined in the proof of
Lemma~\ref{lem:boundary_decay}, and
$\langle\cdot\rangle_{\mathrm{skel}}$
denotes the thermal expectation over
the skeleton spins.

By the definition of $h_k$, we have
\[
|\langle\sigma_v\sigma_w\rangle_N|
\le
h_k(v;\boldsymbol J_C)
h_k(w;\boldsymbol J_D).
\]
Since the couplings in distinct cells
are independent and $0\le h_k\le1$,
we obtain
\begin{align}
\mathbb E[
\langle\sigma_v\sigma_w\rangle_N^2]
&\le
\mathbb E[h_k(v;\boldsymbol J_C)^2]
\mathbb E[h_k(w;\boldsymbol J_D)^2]
\nonumber\\
&\le
\mathbb E[h_k(v;\boldsymbol J_C)]
\mathbb E[h_k(w;\boldsymbol J_D)].
\label{eq:two_point_bound}
\end{align}
Recall that the disorder-averaged squared
site overlap is
\[
\begin{aligned}
R_N(\beta)
&:=
\mathbb E[\langle Q_N^2\rangle_N]\\
&=
\frac{1}{|V^{(N)}|^2}
\sum_{v,w\in V^{(N)}}
\mathbb E[
\langle\sigma_v\sigma_w\rangle_N^2].
\end{aligned}
\]

We divide the sum into three contributions:
pairs of interior vertices belonging to
distinct cells, pairs belonging to the
same cell, and pairs with at least one
skeleton vertex.
For pairs belonging to distinct cells,
Eq.~\eqref{eq:two_point_bound} and the
definition of $B_k$ give
\[
\begin{aligned}
&\frac{1}{|V^{(N)}|^2}
\sum_{\substack{C,D\in\mathcal C_{N,k}\\C\ne D}}
\sum_{\substack{v\in V_{\mathrm{int}}(C)\\
w\in V_{\mathrm{int}}(D)}}
\mathbb E[
\langle\sigma_v\sigma_w\rangle_N^2]
\\
&\qquad\le
\frac{m(m-1)I_k^2B_k^2}{|V^{(N)}|^2}
\le B_k^2.
\end{aligned}
\]

For pairs belonging to the same cell,
we use the trivial bound
$|\langle\sigma_v\sigma_w\rangle_N|\le1$.
Their total contribution is bounded by
\[
\frac{mI_k^2}{|V^{(N)}|^2}
\le\frac{1}{m}
=(bs)^{-(N-k)}.
\]

Similarly, the contribution from pairs
with at least one skeleton vertex is
bounded by
\[
2\frac{|V^{(N-k)}|}{|V^{(N)}|}.
\]

Combining these estimates yields
\begin{equation}
R_N(\beta)
\le
B_k^2+(bs)^{-(N-k)}
+2\frac{|V^{(N-k)}|}{|V^{(N)}|}.
\label{eq:overlap_upper}
\end{equation}
The number of vertices in the
generation-$n$ MK lattice is
\[
|V^{(n)}|
=
2+s(b-1)\frac{(bs)^n-1}{bs-1}.
\]
Hence, for fixed $k$,
\[
\lim_{N\to\infty}
\frac{|V^{(N-k)}|}{|V^{(N)}|}
=(bs)^{-k}.
\]
Taking the limit superior as
$N\to\infty$ in Eq.~\eqref{eq:overlap_upper},
we obtain
\[
\limsup_{N\to\infty}R_N(\beta)
\le B_k^2+2(bs)^{-k}.
\]

Finally, Lemma~\ref{lem:boundary_decay}
implies that $B_k\to0$ as $k\to\infty$
whenever $\Lambda(b,s)\le1$.
Since $bs>1$, letting $k\to\infty$
gives
\[
\lim_{N\to\infty}R_N(\beta)=0,
\]
which completes the proof.
\end{proof}

\subsection{Sufficient condition for  $\Lambda(b,s)\le 1$ }

So far, we have assumed that $\Lambda(b,s)\le1$.
We now derive an explicit sufficient condition for this inequality in terms of the effective dimension $d=1+\log s/\log b$,
thereby completing the proof of Theorem~\ref{thm:sg_absence}.

\begin{lemma}\label{lem:dimension_bound}
For every integer $b\ge2$, define
\begin{equation}
\gamma_b:=
\mathbb E\left[\min_{1\le j\le b}Z_j^2\right]
=
2\int_0^\infty
t\,\operatorname{erfc}(t/\sqrt2)^b\,dt,
\label{eq:gamma}
\end{equation}
where $Z_1,\ldots,Z_b$ are independent
standard Gaussian random variables.
Then
\begin{equation}
\gamma_b\le
\min\left\{
\frac{6}{(b+1)(b+2)},
\frac{\pi}{b^2}
\right\}.
\label{eq:gamma_general_bounds}
\end{equation}
Furthermore, the sharper bound
\begin{equation}
\gamma_b<
\frac{24}{7(b+1)(b+2)}
\qquad(b\ge6)
\label{eq:gamma_refined_bound}
\end{equation}
holds.

For every integer $s\ge 2$, these bounds imply
\begin{equation}
\Lambda(b,s)\le\sqrt{s\gamma_b}.
\label{eq:gammaLambda}
\end{equation}
In particular, if
\begin{equation}
d=1+\frac{\log s}{\log b}
<
1+\frac{\log17}{\log6},
\label{eq:dimension_threshold}
\end{equation}
then $\Lambda(b,s)<1$.
\end{lemma}

\begin{proof}
We first establish Eq.~\eqref{eq:gammaLambda}.
Consider a centered Gaussian scale mixture
$Y=\sqrt{V}\,Z$, where $V>0$ is independent
of $Z\sim\mathcal N(0,1)$ and
$\mathbb E[V]<\infty$.
Let $Y_1,\ldots,Y_b$ be independent copies
of $Y$, represented as $Y_j=\sqrt{V_j}Z_j$.

For fixed $Z_1,\ldots,Z_b$, the function
\[
(v_1,\ldots,v_b)
\longmapsto
\min_{1\le j\le b}v_jZ_j^2
\]
is concave.
Conditioning on $Z_1,\ldots,Z_b$ and
applying Jensen's inequality, we obtain
\[
\begin{aligned}
&\mathbb E\left[
\min_{1\le j\le b}V_jZ_j^2
\mid Z_1,\ldots,Z_b
\right]\\
&\qquad\le
\mathbb E[V]\min_{1\le j\le b}Z_j^2.
\end{aligned}
\]
Taking the expectation over the Gaussian
variables yields
\begin{equation}
\mathbb E\left[\min_{1\le j\le b}Y_j^2\right]
\le\gamma_b\,\mathbb E[Y^2].
\label{eq:gsm-min}
\end{equation}

By Lemma~\ref{lem:gsm}, the
zero-temperature effective coupling
$G_n=(\RG_\infty)^n(G)$ remains a
centered Gaussian scale mixture at
every generation.
Moreover, its second moment is finite,
since the zero-temperature RG recursion
implies
$\mathbb E[G_{n+1}^2]\le s\mathbb E[G_n^2]$.

Since the parallel branches are independent
and have zero mean, Eq.~\eqref{eq:gsm-min}
gives
\[
\begin{aligned}
\mathbb E[G_{n+1}^2]
&=
s\,\mathbb E\left[
\min_{1\le j\le b}G_{n,j}^2
\right]\\
&\le s\gamma_b\,\mathbb E[G_n^2].
\end{aligned}
\]
Together with $\mathbb E[G_0^2]=1$,
this implies
\[
\mathbb E[G_n^2]\le(s\gamma_b)^n.
\]
Furthermore, the peakedness comparison
$\alpha_nG\preceq G_n$, established in
the proof of
Lemma~\ref{lem:supermultiplicativity},
implies
\[
\alpha_n^2
\le\mathbb E[G_n^2]
\le(s\gamma_b)^n.
\]
Taking the $n$-th root and letting
$n\to\infty$, we obtain
\[
\Lambda(b,s)\le\sqrt{s\gamma_b}.
\]

The bounds in
Eqs.~\eqref{eq:gamma_general_bounds}
and~\eqref{eq:gamma_refined_bound}
are established in
Appendix~\ref{app:gamma_bounds}.

It remains to prove the dimension criterion.
By Eq.~\eqref{eq:gammaLambda},
it suffices to show that $s\gamma_b<1$.
Set
\[
a:=\frac{\log17}{\log6}.
\]
The condition in
Eq.~\eqref{eq:dimension_threshold}
is equivalent to $s<b^a$.

For $b\ge10$, consider the function
\[
f(x):=\frac{(x+1)(x+2)}{x^a}.
\]
Since $a<19/12$ and
\[
\frac{x}{x+1}+\frac{x}{x+2}
>\frac{19}{12}>a
\qquad(x\ge10),
\]
the function $f$ is strictly increasing
on $[10,\infty)$.
Moreover,
\[
10^a<10^{19/12}
<\frac{7}{24}(11)(12).
\]
Consequently,
\[
b^a<
\frac{7}{24}(b+1)(b+2)
\qquad(b\ge10).
\]
Using Eq.~\eqref{eq:gamma_refined_bound},
we obtain
\[
s\gamma_b
<
\frac{24b^a}{7(b+1)(b+2)}
<1.
\]

For $2\le b\le9$, the largest admissible
integer values of $s$ are, respectively,
\[
2,\ 5,\ 8,\ 12,\ 16,\ 21,\ 26,\ 32.
\]
These bounds follow from $a<19/12$
and the corresponding integer-power
comparisons, with $6^a=17$.
For $2\le b\le5$, the explicit bounds
\[
\gamma_2<\frac12,\qquad
\gamma_3<\frac15,\qquad
\gamma_4<\frac18,\qquad
\gamma_5<\frac1{12},
\]
established in
Appendix~\ref{app:gamma_bounds},
imply $s\gamma_b<1$.

For $6\le b\le9$, the refined bound
in Eq.~\eqref{eq:gamma_refined_bound}
gives
\[
s\gamma_b
<
\frac{24s}{7(b+1)(b+2)}
\le1.
\]
Hence, $s\gamma_b<1$ whenever
$d<1+\log17/\log6$.
Equation~\eqref{eq:gammaLambda}
then yields
\[
\Lambda(b,s)<1,
\]
which completes the proof.
\end{proof}

Combining Lemmas~\ref{lem:site_overlap} and~\ref{lem:dimension_bound} completes the proof of Theorem~\ref{thm:sg_absence}.

\section{Proof of Theorem~\ref{thm:finite_step}}


In this section, we prove Theorem~\ref{thm:finite_step},
which establishes that, whenever
$\Lambda(b,s)\neq1$, a finite-step procedure
can determine whether $\Lambda(b,s)>1$.

Lemma~\ref{Asymptotic growth rate} provides
lower bounds on $\Lambda(b,s)$ in terms of
$\alpha_n$.
To derive complementary upper bounds,
we first establish a reverse bound to
supermultiplicativity, with a prefactor
that grows linearly in the number of RG steps.

\begin{lemma}
\label{lem:reverse_supermultiplicativity}
Let \(C(b,s)\) be the constant defined in Eq. \eqref{C(b,s)}.
Then, for all integers $n\ge1$ and $m\ge0$,
\begin{equation}
\alpha_{n+m}\le C(b,s) n\,\alpha_n\alpha_m.
\label{eq:reverse_supermultiplicativity}
\end{equation}
\end{lemma}
\begin{proof}
Recall that $g_m$ is the probability density
of the zero-temperature effective coupling
$G_m$, and that
\[
\alpha_m=\frac{\phi(0)}{g_m(0)},
\qquad
c_0=2\phi(0)=\sqrt{\frac{2}{\pi}}.
\]
The proof of Lemma~\ref{lem:tail} gives
\begin{equation}
\mathbb P(|G_m|>t)
\le
\exp\left(-\frac{c_0t}{\alpha_m}\right).
\label{eq:reverse_tail}
\end{equation}
Fix $n\ge1$ and $m\ge0$, and set
\[
K:=(bs)^n,
\qquad
R:=\frac{\alpha_m}{c_0}\log(4K).
\]
By Eq.~\eqref{eq:reverse_tail},
\[
\mathbb P(|G_m|>R)\le\frac{1}{4K}.
\]
The coupling $G_{n+m}$ can be constructed
by applying $n$ zero-temperature RG steps
to $K$ independent copies of $G_m$.
Denote these input couplings by
$Y_1,\ldots,Y_K$, and define
\[
\mathcal A_R:=
\bigcap_{i=1}^K\{|Y_i|\le R\}.
\]
By the union bound,
\[
\mathbb P(\mathcal A_R)
\ge
1-K\mathbb P(|G_m|>R)
\ge\frac34.
\]
Let $X$ have the conditional distribution
\[
X\overset{d}{=}
(G_m\mid |G_m|\le R).
\]
Since the $Y_i$ are independent and
$\mathcal A_R$ is the intersection of
events involving individual inputs,
conditioning on $\mathcal A_R$ preserves
their independence, with each input
distributed as $X$.
Consequently, for every $t\ge0$,
\[
\mathbb P(|G_{n+m}|\le t\mid\mathcal A_R)
=
F_{(\RG_\infty)^n(X)}(t).
\]
Thus, for every $t\ge0$,
\begin{align}
F_{G_{n+m}}(t)
&=\mathbb P(|G_{n+m}|\le t)
\nonumber\\
&\ge
\mathbb P(\mathcal A_R)
\mathbb P(|G_{n+m}|\le t\mid\mathcal A_R)
\nonumber\\
&\ge
\frac34 F_{(\RG_\infty)^n(X)}(t),
\label{eq:reverse_conditioning}
\end{align}
where $F_Y(t):=\mathbb P(|Y|\le t)$.

We next compare $X$ with a rescaled
standard Gaussian random variable.
Since $g_m$ is symmetric unimodal by
Lemma~\ref{preservation-unimodality},
the truncated density of $X$ is also
symmetric unimodal.
Consequently, $F_X$ is concave on
$[0,R]$.
Using $F_X(0)=0$ and $F_X(R)=1$, we obtain
\[
F_X(t)\ge\frac{t}{R}
\qquad(0\le t\le R).
\]
On the other hand, for a standard
Gaussian random variable $G$,
\[
F_G(u)
=
2\int_0^u\phi(x)\,dx
\le c_0u
\qquad(u\ge0).
\]
Therefore, for $0\le t\le R$,
\[
F_G\left(\frac{t}{c_0R}\right)
\le\frac{t}{R}
\le F_X(t).
\]
The same comparison holds for $t\ge R$,
since $F_X(t)=1$.
Hence,
\begin{equation}
X\preceq c_0RG.
\label{eq:reverse_truncated_order}
\end{equation}
By Lemma~\ref{lem:order}, the peakedness
order is preserved under $\RG_\infty$.
Using also the positive homogeneity of
the zero-temperature RG transformation,
we obtain
\[
\begin{aligned}
(\RG_\infty)^n(X)
&\preceq
(\RG_\infty)^n(c_0RG)\\
&\overset{d}{=}c_0RG_n.
\end{aligned}
\]
Equivalently,
\[
F_{(\RG_\infty)^n(X)}(t)
\ge
F_{G_n}\left(\frac{t}{c_0R}\right).
\]
Combining this with
Eq.~\eqref{eq:reverse_conditioning} yields
\begin{equation}
F_{G_{n+m}}(t)
\ge
\frac34
F_{G_n}\left(\frac{t}{c_0R}\right).
\label{eq:reverse_distribution}
\end{equation}

Since $g_n$ and $g_{n+m}$ are symmetric
and continuous at the origin,
\[
\lim_{t\downarrow0}
\frac{F_{G_j}(t)}{2t}
=g_j(0)
\qquad(j=n,n+m).
\]
Dividing Eq.~\eqref{eq:reverse_distribution}
by $2t$ and taking $t\downarrow0$ gives
\[
g_{n+m}(0)
\ge
\frac{3}{4c_0R}g_n(0).
\]
By the definition of $\alpha_n$,
\[
\begin{aligned}
\alpha_{n+m}
&=\frac{\phi(0)}{g_{n+m}(0)}\\
&\le
\frac43c_0R
\frac{\phi(0)}{g_n(0)}\\
&=\frac43c_0R\,\alpha_n\\
&=\frac43\log(4K)\,\alpha_n\alpha_m.
\end{aligned}
\]
Thus, we have established the stronger bound
\begin{equation}
\alpha_{n+m}
\le
\frac43\log\left(4(bs)^n\right)
\alpha_n\alpha_m.
\label{eq:reverse_logarithmic}
\end{equation}
Finally, since $bs\ge2$ and $n\ge1$,
\[
4(bs)^n\le(bs)^{3n}.
\]
Consequently,
\[
\begin{aligned}
\frac43\log\left(4(bs)^n\right)
&\le4n\log(bs)\\
&\le4n\lceil\log_2(bs)\rceil\\
&=C(b,s)n.
\end{aligned}
\]
Combining this with
Eq.~\eqref{eq:reverse_logarithmic},
we conclude that
\[
\alpha_{n+m}\le C(b,s) n\,\alpha_n\alpha_m,
\]
which completes the proof.
\end{proof}

By Lemma~\ref{lem:reverse_supermultiplicativity},
we obtain the following upper bound on $\Lambda(b,s)$.

\begin{lemma}
\label{lem:lambda_upper}
For every integer $n\ge1$,
\begin{equation}
\Lambda(b,s)^n\le C(b,s) n\,\alpha_n.
\end{equation}
In particular, if $C(b,s)n\,\alpha_n<1$
for some $n\ge1$, then $\Lambda(b,s)<1$.
\end{lemma}
\begin{proof}
Iterating the inequality in
Lemma~\ref{lem:reverse_supermultiplicativity}
with a fixed $n\ge1$ gives
\[
\alpha_{rn}
\le(C(b,s)n)^{r-1}\alpha_n^r
\]
for every integer $r\ge1$.
Taking the $r$-th root yields
\[
\alpha_{rn}^{1/r}
\le(C(b,s)n)^{1-1/r}\alpha_n.
\]
By Lemma~\ref{Asymptotic growth rate},
\[
\lim_{r\to\infty}\alpha_{rn}^{1/r}
=
\Lambda(b,s)^n.
\]
Hence, letting $r\to\infty$, we obtain
\[
\Lambda(b,s)^n\le C(b,s)n\,\alpha_n.
\]
The final assertion follows immediately.
\end{proof}

We are now ready to prove
Theorem~\ref{thm:finite_step}.

\begin{proof}[Proof of Theorem~\ref{thm:finite_step}]
Combining Lemmas~\ref{Asymptotic growth rate}
and~\ref{lem:lambda_upper}, we obtain
\begin{equation}
\alpha_n
\le\Lambda(b,s)^n
\le C(b,s)n\,\alpha_n
\label{eq:lambda_two_sided}
\end{equation}
for every integer $n\ge1$, where
$C(b,s)=4\lceil\log_2(bs)\rceil$.
These inequalities immediately imply
\[
\alpha_n>1
\quad\Longrightarrow\quad
\Lambda(b,s)>1,
\]
and
\[
C(b,s)n\,\alpha_n<1
\quad\Longrightarrow\quad
\Lambda(b,s)<1.
\]
We now show that at least one of these
two stopping criteria is satisfied after
finitely many RG steps whenever
$\Lambda(b,s)\neq1$.

First, suppose that $\Lambda(b,s)>1$.
If $n$ satisfies
\[
\Lambda(b,s)^n>C(b,s)n,
\]
then Eq.~\eqref{eq:lambda_two_sided} gives
\[
\alpha_n
\ge\frac{\Lambda(b,s)^n}{C(b,s)n}>1.
\]
Thus, the first stopping criterion is satisfied.

Next, suppose that $\Lambda(b,s)<1$.
If $n$ satisfies
\[
\Lambda(b,s)^{-n}>C(b,s)n,
\]
then Eq.~\eqref{eq:lambda_two_sided} yields
\[
C(b,s)n\,\alpha_n
\le C(b,s)n\,\Lambda(b,s)^n<1.
\]
Thus, the second stopping criterion is satisfied.

In either case, a sufficient condition
for termination is
\begin{equation}
\exp\left(n|\log\Lambda(b,s)|\right)>C(b,s)n.
\label{eq:termination_condition}
\end{equation}
Since $\Lambda(b,s)\neq1$, the left-hand
side grows exponentially with $n$,
whereas the right-hand side grows only
linearly.
Therefore, Eq.~\eqref{eq:termination_condition}
holds for all sufficiently large $n$.

Hence, the procedure terminates after
finitely many RG steps whenever
$\Lambda(b,s)\neq1$, completing the proof.
\end{proof}

\section{Proof of Theorem~\ref{thm:sg_existence}}

\subsection{Reference spin configuration and Peierls-type argument}
We now prove the existence of finite-temperature
spin-glass order.
Our approach is based on a Peierls-type argument
that bounds thermal deviations from a
disorder- and temperature-dependent reference
spin configuration, defined below.

We construct the reference configuration
recursively using the finite-temperature
effective couplings.
Consider a generation-$(k+1)$ parent cell
consisting of $s$ parallel branches, each
containing $b$ generation-$k$ child cells
connected in series.
Tracing out the internal spins of each child
cell while retaining its two endpoint spins
yields effective couplings $K_{i,j}^{(k)}$,
where $1\le i\le s$ labels the branches
and $1\le j\le b$ labels the positions
along each branch.
These couplings are independent copies of
$J_k(\beta)$, so that
\[
K_{i,j}^{(k)}\overset{d}{=}J_k(\beta).
\]

Up to an additive constant independent
of the skeleton spins, the resulting
effective Hamiltonian is
\[
H_{\mathrm{eff}}^{(k)}
=
-\sum_{i=1}^s\sum_{j=1}^b
K_{i,j}^{(k)}
\sigma_{i,j-1}\sigma_{i,j},
\]
where
\[
\sigma_{i,0}=\sigma_L,
\qquad
\sigma_{i,b}=\sigma_R
\]
are the boundary spins of the parent cell.

For fixed boundary spins, we choose the
internal skeleton spins to minimize
$H_{\mathrm{eff}}^{(k)}$.
Since the branches share only their
boundary spins, this minimization can
be performed independently on each branch.
If the prescribed boundary spins are
compatible with satisfying all effective
bonds along a branch, all bonds are satisfied.
Otherwise, the minimizing configuration
leaves exactly one bond unsatisfied,
namely the bond with the smallest
absolute effective coupling.
Since the microscopic couplings have
continuous distributions, degeneracies
in this minimization occur with probability zero.

For the generation-$N$ MK lattice,
we first fix the left boundary spin
of the reference configuration to
$\sigma_A^{\mathrm{ref}}=+1$ and choose
\[
\sigma_B^{\mathrm{ref}}
=
\operatorname{sgn}J_N(\beta).
\]
This choice minimizes the effective
interaction between the two boundary spins.
We then recursively apply the above
minimization procedure to each cell,
proceeding from larger to smaller
generations.
This determines a reference configuration
$\{\sigma_v^{\mathrm{ref}}\}_{v\in V^{(N)}}$
on the entire lattice.

The reference configuration depends on
both the disorder realization and the
inverse temperature $\beta$.
At finite temperature, the effective
couplings incorporate entropic contributions
from the internal degrees of freedom.
Consequently, the reference configuration
need not be a ground state of the original
Hamiltonian $H_N$.

At zero temperature, by contrast,
the effective couplings describe the
energy differences between fixed boundary
conditions, and the RG transformation
reduces to exact energy minimization.
The same recursive construction therefore
yields a ground state of $H_N$,
up to a global spin flip.

The fixing of $\sigma_A^{\mathrm{ref}}=+1$
serves only to remove the global spin-flip
ambiguity of the reference configuration;
the Gibbs measure retains free boundary
conditions.

We next develop a Peierls-type argument
to bound thermal deviations from the
reference configuration, modulo a global
spin flip, and use these estimates to
derive a lower bound on the spin-glass order parameter.

\begin{lemma}
\label{lem:peierls}
Let $p_{n,\beta}$ denote the probability density
of the generation-$n$ effective coupling
$J_n(\beta)$, and define
\[
D_n(\beta):=\|p_{n,\beta}\|_\infty.
\]
For the generation-$N$ MK lattice, let
\begin{equation}
\delta_N(\beta)
:=
\frac{\log 2}{\beta}D_N(\beta)
+
\frac{sb(b-1)}{\beta}
\sum_{n=0}^{N-1}D_n(\beta).
\label{eq:peierls_delta}
\end{equation}
Then
\begin{equation}
\mathbb E[\langle Q_N^2\rangle_N]
\ge
\left[
\max\{0,1-4\delta_N(\beta)\}
\right]^2.
\label{eq:peierls_overlap}
\end{equation}
\end{lemma}

\begin{proof}
Fix $0<\beta<\infty$. If $\delta_N(\beta)=\infty$,
the assertion is trivial; hence we assume
$\delta_N(\beta)<\infty$.
For any event $E$ involving spins in the generation-$N$
lattice, write
\[
\mu_N(E):=\langle\mathbf 1_E\rangle_N.
\]
The proof consists of three steps.

\medskip
\noindent

\textit{Step 1: Thermal deviations within a single cell.}

Consider a generation-$(n+1)$ cell $C$, where $0\le n<N$.
It consists of $s$ parallel branches, each containing
$b$ generation-$n$ child cells in series.
After tracing out the internal spins of these child cells,
the effective Hamiltonian is, up to a constant independent
of the skeleton spins,
\[
H_{\mathrm{eff}}^{(n)}
=-\sum_{i=1}^{s}\sum_{j=1}^{b}
K_{i,j}^{(n)}\sigma_{i,j-1}\sigma_{i,j},
\]
where the $K_{i,j}^{(n)}$ are independent copies of
$J_n(\beta)$.

Fix the two endpoint spins of $C$ and consider one branch.
Suppressing its index, write the effective couplings as
$K_1,\ldots,K_b$ and the skeleton spins as
$\sigma_0,\ldots,\sigma_b$.
Let $\widehat{\sigma}_0,\ldots,\widehat{\sigma}_b$
denote a configuration minimizing the branch effective
Hamiltonian for these fixed endpoints.
If all effective bonds can be satisfied,
the minimizer satisfies every bond. Otherwise, exactly
one bond with the smallest absolute coupling is left
unsatisfied. Since the effective couplings have continuous
distributions, the weakest bond is unique almost surely.
Set
\[
r:=\operatorname*{arg\,min}_{1\le j\le b}|K_j|.
\]
Starting from the left endpoint, determine the minimizing
spins through $\widehat{\sigma}_{r-1}$ by satisfying
successive bonds. Starting from the right endpoint,
determine the minimizing spins through
$\widehat{\sigma}_r$ in the same way.
This produces the required minimizing configuration.

Define
\[
w_j:=K_j\widehat{\sigma}_{j-1}
\widehat{\sigma}_j,
\qquad
\lambda_j:=\sigma_j\widehat{\sigma}_j.
\]
By construction, $w_j=|K_j|$ for $j\ne r$, whereas
$w_r=\pm|K_r|$. Since the endpoints are fixed,
$\lambda_0=\lambda_b=1$. Thus, up to an additive constant,
\[
H(\boldsymbol\lambda)
=-\sum_{j=1}^{b}w_j\lambda_{j-1}\lambda_j.
\]

For $1\le a<c\le b$, let $A_{a,c}$ be the event that
$\{a,\ldots,c-1\}$ is a maximal interval of negative
variables, namely,
\[
\lambda_{a-1}=\lambda_c=1,
\qquad
\lambda_j=-1\quad(a\le j<c).
\]
Let $f_{a,c}\boldsymbol\lambda$ be the configuration
obtained by changing the spins in this interval from
$-1$ to $+1$ and leaving all other spins unchanged.
Only the two boundary bonds of the interval change,
so that
\[
H(\boldsymbol\lambda)
-H(f_{a,c}\boldsymbol\lambda)=2(w_a+w_c).
\]
At most one $w_j$ is negative, and if it is negative,
then $|w_j|$ is the smallest absolute coupling.
Consequently,
\[
w_a+w_c\ge\bigl||K_a|-|K_c|\bigr|,
\]
and hence
\begin{equation}
H(\boldsymbol\lambda)-H(f_{a,c}\boldsymbol\lambda)
\ge2\bigl||K_a|-|K_c|\bigr|.
\label{eq:peierls_energy}
\end{equation}

The map $f_{a,c}$ is injective on $A_{a,c}$.
Therefore,
\[
\begin{aligned}
&\sum_{\boldsymbol\lambda\in A_{a,c}}
 e^{-\beta H(\boldsymbol\lambda)}\\
&\quad\le
 e^{-2\beta\left||K_a|-|K_c|\right|}
 \sum_{\boldsymbol\lambda\in A_{a,c}}
 e^{-\beta H(f_{a,c}\boldsymbol\lambda)}\\
&\quad\le
 e^{-2\beta\left||K_a|-|K_c|\right|}Z_{\mathrm{br}},
\end{aligned}
\]
where $Z_{\mathrm{br}}$ is the branch partition function
with its endpoint spins and effective couplings fixed.
Let $\mu_{\mathrm{br}}$ denote the corresponding Gibbs
probability measure. Then
\begin{equation}
\begin{aligned}
\mu_{\mathrm{br}}(A_{a,c})
&=\frac1{Z_{\mathrm{br}}}
\sum_{\boldsymbol\lambda\in A_{a,c}}
 e^{-\beta H(\boldsymbol\lambda)}\\
&\le e^{-2\beta\left||K_a|-|K_c|\right|}.
\end{aligned}
\label{eq:peierls_interval}
\end{equation}

Let $E_C$ be the event that at least one internal
skeleton spin of $C$ differs from the minimizing
configuration for the actual endpoint spins of $C$.
Let $\mu_C(\,\cdot\mid\sigma_L,\sigma_R)$ denote
the Gibbs probability measure of the skeleton spins
of $C$, with its two endpoint spins and effective
couplings fixed. Conditional on the endpoints, its
branches are independent. Every occurrence of $E_C$
contains a maximal interval of negative variables in
some branch. Thus, by the union bound and
Eq.~\eqref{eq:peierls_interval},
\begin{equation}
\begin{aligned}
&\mu_C(E_C\mid\sigma_L,\sigma_R)\\
&\quad\le
\sum_{i=1}^{s}\sum_{1\le a<c\le b}
 e^{-2\beta\left|
 |K_{i,a}^{(n)}|-|K_{i,c}^{(n)}|
 \right|}.
\end{aligned}
\label{eq:peierls_cell}
\end{equation}
This estimate holds for every choice of endpoint spins
and every realization of the effective couplings.

We now average its right-hand side over the disorder.
The variables $K_{i,a}^{(n)}$ and $K_{i,c}^{(n)}$ are
independent and have the common symmetric density
$p_{n,\beta}$. Hence their absolute values have density
\[
f_n(x)=2p_{n,\beta}(x),\qquad x>0.
\]
Since $f_n(x)\le2D_n(\beta)$, we obtain
\[
\begin{aligned}
&\mathbb E\left[
 e^{-2\beta\left||K_{i,a}^{(n)}|-|K_{i,c}^{(n)}|\right|}
 \right]\\
&=\int_0^\infty f_n(y)
 \int_0^\infty f_n(x)e^{-2\beta|x-y|}\,dx\,dy\\
&\le2D_n(\beta)
 \int_0^\infty f_n(y)
 \int_0^\infty e^{-2\beta|x-y|}\,dx\,dy\\
&\le\frac{2D_n(\beta)}{\beta},
\end{aligned}
\]
where we used
\[
\int_0^\infty e^{-2\beta|x-y|}\,dx\le\frac1\beta
\qquad(y\ge0).
\]
There are $s\binom b2$ choices of a branch and interval
endpoints. Therefore,
\begin{equation}
\begin{aligned}
&\mathbb E\left[
 \sum_{i=1}^{s}\sum_{1\le a<c\le b}
 e^{-2\beta\left|
 |K_{i,a}^{(n)}|-|K_{i,c}^{(n)}|
 \right|}\right]\\
&\quad\le
 \frac{sb(b-1)}{\beta}D_n(\beta).
\end{aligned}
\label{eq:peierls_cell_average}
\end{equation}

When $C$ is embedded in the generation-$N$ lattice,
its internal spins are conditionally independent of the
rest of the lattice given its endpoint spins, for fixed
disorder. Because Eq.~\eqref{eq:peierls_cell} holds
uniformly over the endpoint spins, it follows that
\[
\mu_N(E_C)\le
\sum_{i=1}^{s}\sum_{1\le a<c\le b}
 e^{-2\beta\left|
 |K_{i,a}^{(n)}|-|K_{i,c}^{(n)}|
 \right|}.
\]
Taking the disorder average and using
Eq.~\eqref{eq:peierls_cell_average}, we obtain
\begin{equation}
\mathbb E[\mu_N(E_C)]
\le\frac{sb(b-1)}{\beta}D_n(\beta).
\label{eq:peierls_cell_global}
\end{equation}
In particular, no independence between the endpoint
spins and the effective couplings is assumed.

\medskip
\noindent

\textit{Step 2: Construction of global reference configuration.}

Let $A,B$ denote the two outer boundary vertices.
Construct a reference configuration
$\sigma_v^{\mathrm{ref}}\in\{-1,+1\}$
for $v\in V^{(N)}$ by setting
\[
\sigma_A^{\mathrm{ref}}=1,\qquad
\sigma_B^{\mathrm{ref}}=\operatorname{sgn}J_N(\beta),
\]
and recursively minimizing the effective Hamiltonian
in every cell, from the highest generation to the lowest.
The reference configuration depends only on the disorder
and $\beta$, not on the thermal spin configuration.
To respect the global spin-flip symmetry of the free
boundary Gibbs measure, we compare $\sigma_v$ with
$\sigma_v^{\mathrm{ref}}\sigma_A$.

Define the top-level mismatch event
\[
E_*:=\{\sigma_A\sigma_B
\ne\operatorname{sgn}J_N(\beta)\}.
\]
Tracing out the internal spins gives the effective
two-spin Hamiltonian
$H_{\mathrm{top}}=-J_N(\beta)\sigma_A\sigma_B$.
Hence, for fixed disorder,
\[
\mu_N(E_*)=
\frac1{1+e^{2\beta|J_N(\beta)|}}.
\]
By the symmetry of $p_{N,\beta}$,
\[
\begin{aligned}
\mathbb E[\mu_N(E_*)]
&=2\int_0^\infty
\frac{p_{N,\beta}(x)}{1+e^{2\beta x}}\,dx\\
&\le2D_N(\beta)\int_0^\infty
\frac{dx}{1+e^{2\beta x}}\\
&=\frac{\log2}{\beta}D_N(\beta).
\end{aligned}
\]
Thus,
\begin{equation}
\mathbb E[\mu_N(E_*)]
\le\frac{\log2}{\beta}D_N(\beta).
\label{eq:peierls_top}
\end{equation}

Fix $v\in V^{(N)}\setminus\{A,B\}$.
Starting from the generation-$N$ cell,
recursively descend into the unique child
cell whose interior contains $v$, until
$v$ is an internal skeleton vertex of
the current cell.
Let $\mathcal A_N(v)$ denote the sequence
of cells visited in this procedure,
including the terminal cell.
For $v=A$ or $v=B$, set
$\mathcal A_N(v)=\varnothing$.

Since the interiors of distinct child cells
are disjoint, this sequence is uniquely
determined by the hierarchical geometry
and contains at most one cell of each
generation.

If $E_*$ does not occur and none of
the events $E_C$ occurs along this
sequence, the spins agree with the
global reference configuration, up to
the global spin flip $\sigma_A$.
In particular,
$\sigma_v=\sigma_v^{\mathrm{ref}}\sigma_A$.
Therefore,
\begin{equation}
\{\sigma_v\ne\sigma_v^{\mathrm{ref}}\sigma_A\}
\subseteq
E_*\cup\bigcup_{C\in\mathcal A_N(v)}E_C.
\label{eq:reference_mismatch}
\end{equation}
By the union bound and
Eqs.~\eqref{eq:peierls_cell_global}
and~\eqref{eq:peierls_top},
\[
\begin{aligned}
&\mathbb E[
\mu_N(\sigma_v\ne\sigma_v^{\mathrm{ref}}\sigma_A)]\\
&\quad\le
\frac{\log2}{\beta}D_N(\beta)
+\frac{sb(b-1)}{\beta}
\sum_{n=0}^{N-1}D_n(\beta)\\
&\quad=\delta_N(\beta).
\end{aligned}
\]
Thus, uniformly over $v\in V^{(N)}$,
\begin{equation}
\mathbb E[
\mu_N(\sigma_v\ne\sigma_v^{\mathrm{ref}}\sigma_A)]
\le\delta_N(\beta).
\label{eq:peierls_vertex}
\end{equation}

\medskip
\noindent
\textit{Step 3: Lower bound on spin-glass order parameter.}

Let $u,v\in V^{(N)}$. If both spins agree with
the global reference configuration up to the
common spin flip $\sigma_A$, then
$\sigma_u\sigma_v
=\sigma_u^{\mathrm{ref}}\sigma_v^{\mathrm{ref}}$.
Therefore,
\[
\begin{aligned}
&\{\sigma_u^{\mathrm{ref}}\sigma_v^{\mathrm{ref}}
\sigma_u\sigma_v=-1\}\\
&\quad\subseteq
\{\sigma_u\ne\sigma_u^{\mathrm{ref}}\sigma_A\}
\cup
\{\sigma_v\ne\sigma_v^{\mathrm{ref}}\sigma_A\}.
\end{aligned}
\]
Using Eq.~\eqref{eq:peierls_vertex}, we obtain
\begin{equation}
\begin{aligned}
&\mathbb E\bigl[
\mu_N(
\sigma_u^{\mathrm{ref}}\sigma_v^{\mathrm{ref}}
\sigma_u\sigma_v=-1)
\bigr]\\
&\quad\le2\delta_N(\beta).
\end{aligned}
\label{eq:peierls_pair_error}
\end{equation}
Since
$\sigma_u^{\mathrm{ref}}\sigma_v^{\mathrm{ref}}
\sigma_u\sigma_v\in\{-1,+1\}$,
\[
\begin{aligned}
&\mathbb E\left[
\sigma_u^{\mathrm{ref}}\sigma_v^{\mathrm{ref}}
\langle\sigma_u\sigma_v\rangle_N
\right]\\
&\quad=1-2\mathbb E\bigl[
\mu_N(
\sigma_u^{\mathrm{ref}}\sigma_v^{\mathrm{ref}}
\sigma_u\sigma_v=-1)
\bigr]\\
&\quad\ge1-4\delta_N(\beta).
\end{aligned}
\]
By Jensen's inequality and
$(\sigma_u^{\mathrm{ref}}
\sigma_v^{\mathrm{ref}})^2=1$,
\[
\begin{aligned}
&\mathbb E[
\langle\sigma_u\sigma_v\rangle_N^2]\\
&=
\mathbb E\left[
\left(
\sigma_u^{\mathrm{ref}}\sigma_v^{\mathrm{ref}}
\langle\sigma_u\sigma_v\rangle_N
\right)^2
\right]\\
&\ge
\left(
\mathbb E\left[
\sigma_u^{\mathrm{ref}}\sigma_v^{\mathrm{ref}}
\langle\sigma_u\sigma_v\rangle_N
\right]
\right)^2.
\end{aligned}
\]
When $1-4\delta_N(\beta)>0$, the preceding
inequalities give a lower bound
$(1-4\delta_N(\beta))^2$; otherwise the trivial
lower bound zero applies. Consequently,
\begin{equation}
\mathbb E[\langle\sigma_u\sigma_v\rangle_N^2]
\ge\left[\max\{0,1-4\delta_N(\beta)\}\right]^2.
\label{eq:peierls_correlation}
\end{equation}
Finally, summing over $u,v\in V^{(N)}$ and using
the definition of $Q_N$, we obtain
\[
\mathbb E[\langle Q_N^2\rangle_N]
\ge\left[\max\{0,1-4\delta_N(\beta)\}\right]^2.
\]
This proves Eq.~\eqref{eq:peierls_overlap}.
\end{proof}

To establish the existence of finite-temperature
spin-glass order, it remains to derive suitable
upper bounds on
$D_n(\beta)=\|p_{n,\beta}\|_\infty$.
In particular, Lemma~\ref{lem:peierls} shows
that the condition
\[
\sup_{N\ge1}\delta_N(\beta)<\frac14
\]
guarantees a strictly positive lower bound
on the spin-glass order parameter,
uniformly in $N$.

\subsection{Existence of finite-temperature spin-glass phase}
We retain the notation
\[
J_n(\beta)=(\RG_\beta)^n(G),
\qquad
G_n=(\RG_\infty)^n(G),
\qquad G\sim\mathcal N(0,1),
\]
with densities $p_{n,\beta}$ and $g_n$, respectively.

To bound $D_n(\beta)$ at sufficiently low temperatures,
we first estimate the difference between the
finite- and zero-temperature RG flows.


\begin{lemma}
\label{lem:finite_zero_convergence}
For every fixed integer $r\ge0$,
\begin{align}
\sup_{\beta\ge1}D_r(\beta)&<\infty,
\label{eq:direct_uniform_density}\\
\lim_{\beta\to\infty}
\|p_{r,\beta}-g_r\|_\infty&=0.
\label{eq:finite_zero_density}
\end{align}
Moreover, when the two flows are constructed from
the same microscopic Gaussian couplings,
\begin{equation}
|J_r(\beta)-G_r|
\le\varepsilon_r(\beta)
:=
\frac{s\log b}{2\beta}
\frac{s^r-1}{s-1}
\qquad\text{almost surely}.
\label{eq:direct_coupling_error}
\end{equation}
\end{lemma}
\begin{proof}
The estimates are established in
Appendix~\ref{app:finite_zero_density}.
\end{proof}

We then show that, when $\Lambda(b,s)>1$,
the finite-temperature effective couplings
grow under successive RG transformations.

\begin{lemma}
\label{lem:finite_expansion}
Suppose that $\Lambda(b,s)>1$.
Then there exist an integer $m\ge1$,
a constant $\lambda>1$, and a finite
$\beta_0\ge1$ such that
\begin{equation}
J_m(\beta)\succeq\lambda G
\qquad(\beta\ge\beta_0).
\label{eq:finite_expansion}
\end{equation}
Furthermore, for every integer $k\ge0$,
\begin{equation}
J_{km}(\beta)\succeq\lambda^kG
\qquad(\beta\ge\beta_0).
\label{eq:iterated_expansion}
\end{equation}
\end{lemma}
\begin{proof}
By Lemma~\ref{Asymptotic growth rate},
choose $m\ge1$ such that $\alpha_m>1$,
and fix $1<\lambda<\alpha_m$.
The comparison established in the proof of
Lemma~\ref{lem:supermultiplicativity} gives
\[
G_m\succeq\alpha_mG,
\qquad
g_m(0)=\frac{\phi(0)}{\alpha_m}.
\]
We prove Eq.~\eqref{eq:finite_expansion}
by treating a neighborhood of the origin
and its complement separately.

Since $\lambda<\alpha_m$ and
\[
\lim_{t\downarrow0}
\frac{F_G(t/\lambda)}{2t}
=
\frac{\phi(0)}{\lambda}>g_m(0),
\]
we may choose $t_0>0$ such that
\begin{equation}
\frac{F_G(t_0/\lambda)}{2t_0}>g_m(0).
\label{eq:direct_origin_gap}
\end{equation}
By Eq.~\eqref{eq:finite_zero_density}
and the symmetric unimodality of $g_m$,
we have
\[
D_m(\beta)
\longrightarrow
\|g_m\|_\infty=g_m(0).
\]
Thus, for all sufficiently large $\beta$,
\[
D_m(\beta)
\le\frac{F_G(t_0/\lambda)}{2t_0}.
\]
For $0\le t\le t_0$, the concavity of
$F_G$ on $[0,\infty)$ therefore yields
\begin{equation}
\begin{aligned}
F_{J_m(\beta)}(t)
&\le2tD_m(\beta)\\
&\le\frac{t}{t_0}F_G(t_0/\lambda)\\
&\le F_G(t/\lambda).
\end{aligned}
\label{eq:direct_small_t}
\end{equation}

For $t\ge t_0$,
Eq.~\eqref{eq:direct_coupling_error}
and $G_m\succeq\alpha_mG$ give
\[
\begin{aligned}
F_{J_m(\beta)}(t)
&\le F_{G_m}(t+\varepsilon_m(\beta))\\
&\le F_G\left(
\frac{t+\varepsilon_m(\beta)}{\alpha_m}
\right).
\end{aligned}
\]
Increase the temperature threshold,
if necessary, so that
\[
\varepsilon_m(\beta)
\le
\left(\frac{\alpha_m}{\lambda}-1\right)t_0.
\]
Then
\begin{equation}
F_{J_m(\beta)}(t)\le F_G(t/\lambda)
\qquad(t\ge t_0).
\label{eq:direct_large_t}
\end{equation}
Choose $\beta_0\ge1$ so that both estimates
hold for every $\beta\ge\beta_0$.
Together, Eqs.~\eqref{eq:direct_small_t}
and~\eqref{eq:direct_large_t} establish
Eq.~\eqref{eq:finite_expansion}.

To iterate this comparison, recall
the scaling identity
\begin{equation}
(\RG_\beta)^r(cY)
\overset{d}{=}
c(\RG_{c\beta})^r(Y)
\qquad(c>0,\ r\ge0),
\label{eq:rg_scaling}
\end{equation}
which follows from
\[
\Phi_\beta(cx_1,\ldots,cx_b)
=
c\Phi_{c\beta}(x_1,\ldots,x_b).
\]
The assertion in
Eq.~\eqref{eq:iterated_expansion}
holds for $k=0$, since
$J_0(\beta)\overset{d}{=}G$.
If it holds for some $k\ge0$,
preservation of the peakedness order
and symmetric unimodality gives
\[
\begin{aligned}
J_{(k+1)m}(\beta)
&\overset{d}{=}
(\RG_\beta)^m(J_{km}(\beta))\\
&\succeq
(\RG_\beta)^m(\lambda^kG)\\
&\overset{d}{=}
\lambda^kJ_m(\lambda^k\beta)\\
&\succeq\lambda^{k+1}G.
\end{aligned}
\]
The last inequality follows from
Eq.~\eqref{eq:finite_expansion},
since $\lambda^k\beta\ge\beta_0$.
Induction completes the proof.
\end{proof}

This growth implies exponential decay of
the maximum density $D_n(\beta)$ with
the RG generation, ensuring its summability.

\begin{lemma}
\label{lem:density_decay}
Let $m$, $\lambda$, and $\beta_0$
be chosen as in
Lemma~\ref{lem:finite_expansion}.
For $0\le l<m$, define
\begin{equation}
M_l:=
\sup_{\beta\ge\beta_0}D_l(\beta).
\label{eq:M_l}
\end{equation}
Then $M_l<\infty$, and, for all $k\ge0$,
$0\le l<m$, and $\beta\ge\beta_0$,
\begin{equation}
D_{km+l}(\beta)
\le
\lambda^{-k}D_l(\lambda^k\beta)
\le
\lambda^{-k}M_l.
\label{eq:density_decay}
\end{equation}
Consequently,
\begin{equation}
\sum_{n=0}^{\infty}D_n(\beta)
\le
\frac{1}
{1-\lambda^{-1}}\sum_{l=0}^{m-1}M_l <\infty
\qquad(\beta\ge\beta_0).
\label{eq:density_sum}
\end{equation}
\end{lemma}
\begin{proof}
The finiteness of $M_l$ follows from
Eq.~\eqref{eq:direct_uniform_density}.
By Lemma~\ref{lem:finite_expansion},
preservation of the peakedness order,
and Eq.~\eqref{eq:rg_scaling},
\[
\begin{aligned}
J_{km+l}(\beta)
&\overset{d}{=}
(\RG_\beta)^l(J_{km}(\beta))\\
&\succeq
(\RG_\beta)^l(\lambda^kG)\\
&\overset{d}{=}
\lambda^kJ_l(\lambda^k\beta).
\end{aligned}
\]
Hence, for every $t\ge0$,
\[
\begin{aligned}
F_{J_{km+l}(\beta)}(t)
&\le
F_{J_l(\lambda^k\beta)}(t/\lambda^k)\\
&\le
2t\lambda^{-k}D_l(\lambda^k\beta).
\end{aligned}
\]
Choose the even, nonincreasing representative
of $p_{km+l,\beta}$ on the positive half-line.
For $t>0$,
\[
2t\,p_{km+l,\beta}(t)
\le F_{J_{km+l}(\beta)}(t).
\]
Taking the essential supremum therefore
proves the first inequality in
Eq.~\eqref{eq:density_decay}.
The second follows from
$\lambda^k\beta\ge\beta_0$.
Finally, summing over $n=km+l$,
with $0\le l<m$, gives
Eq.~\eqref{eq:density_sum}.
\end{proof}

combining these density bounds
with the Peierls-type estimate of
Lemma~\ref{lem:peierls},
we establish the existence of
finite-temperature spin-glass order.

\begin{lemma}[Existence of finite-temperature spin-glass order]
\label{lem:finite_temperature_sg}
Suppose that $\Lambda(b,s)>1$.
Then there exists
$0<\beta_{\mathrm{SG}}<\infty$ such that
\begin{equation}
\liminf_{N\to\infty}
\mathbb E[\langle Q_N^2\rangle_N]>0
\qquad(\beta>\beta_{\mathrm{SG}}).
\label{eq:finite_temperature_sg}
\end{equation}
\end{lemma}
\begin{proof}
With $m$, $\lambda$, $\beta_0$,
and $M_l$ as above, define
\[
C_\delta:=
(\log2)\max_{0\le l<m}M_l
+
\frac{sb(b-1)}{1-\lambda^{-1}}
\sum_{l=0}^{m-1}M_l<\infty.
\]
This constant is independent of
$N$ and $\beta$.
By Lemma~\ref{lem:density_decay}
and Eq.~\eqref{eq:peierls_delta},
\begin{equation}
\sup_{N\ge1}\delta_N(\beta)
\le\frac{C_\delta}{\beta}
\qquad(\beta\ge\beta_0).
\label{eq:delta_uniform_bound}
\end{equation}
Set
\[
\beta_{\mathrm{SG}}
:=
\max\{\beta_0,4C_\delta\}.
\]
For every $\beta>\beta_{\mathrm{SG}}$,
Lemma~\ref{lem:peierls} yields
\[
\mathbb E[\langle Q_N^2\rangle_N]
\ge
\left(1-\frac{4C_\delta}{\beta}\right)^2
>0
\qquad(N\ge1).
\]
Taking the limit inferior as
$N\to\infty$ proves the claim.
\end{proof}

\subsection{Sufficient condition for  $\Lambda(b,s)>1$ }

Finally, we derive an explicit sufficient
condition for $\Lambda(b,s)>1$ by comparing
the zero-temperature RG flow with the
uniform distribution.

\begin{lemma}[Explicit lower bound on $\Lambda(b,s)$]
\label{lem:gamma_condition}
For integers $b,s\ge2$, define
\begin{equation}
\Gamma_{b,s}:=
\frac{b}{\sqrt{s}}
\left(
\frac{b\sqrt{s(s-1)}}{bs-1}
\right)^{s-1}.
\label{eq:gamma_bs}
\end{equation}
Then
\begin{equation}
\Lambda(b,s)\ge\Gamma_{b,s}^{-1}.
\label{eq:lambda_gamma_bound}
\end{equation}
In particular, $\Gamma_{b,s}<1$
implies $\Lambda(b,s)>1$.
Moreover, $\Gamma_{b,s}<1$ holds
for all $b,s\ge2$ with fractal dimension
$d=1+\log s/\log b\ge3$.
\end{lemma}
\begin{proof}
We compare the zero-temperature RG flow
with the uniform distribution
$U\sim\mathrm{Unif}[-1,1]$.

The signed minimum of $b$ independent
copies of $U$ has probability density
\[
q_b(x)=\frac{b}{2}(1-|x|)_+^{b-1},
\]
where $(x)_+:=\max\{x,0\}$.
Since the parallel update consists
of the sum of $s$ independent
series-branch outputs, the density
of $\RG_\infty(U)$ is
\[
r_{b,s}=q_b^{*s}.
\]

We estimate its supremum norm using
the sharp Young convolution inequality~\cite{brascamp1976best}.
Set
\[
p:=\frac{s}{s-1},
\qquad p'=s,
\]
so that $s/p=s-1$.
The one-dimensional sharp Young
inequality gives
\begin{equation}
\|r_{b,s}\|_\infty
\le A_p^s\|q_b\|_p^s,
\label{eq:sharp_young_rg}
\end{equation}
where
\[
A_p:=
\left(
\frac{p^{1/p}}{(p')^{1/p'}}
\right)^{1/2}.
\]

A direct calculation yields
\[
\begin{aligned}
\|q_b\|_p^p
&=
2\left(\frac b2\right)^p
\int_0^1(1-x)^{p(b-1)}\,dx\\
&=
\frac{2(b/2)^p}{p(b-1)+1}.
\end{aligned}
\]
Since
\[
p(b-1)+1=\frac{bs-1}{s-1},
\]
we obtain
\[
\|q_b\|_p^s
=
\left(\frac b2\right)^s
\left(
\frac{2(s-1)}{bs-1}
\right)^{s-1}.
\]
Furthermore,
\[
A_p^s
=
\frac{1}{\sqrt{s}}
\left(\frac{s}{s-1}\right)^{(s-1)/2}.
\]
Substituting these expressions into
Eq.~\eqref{eq:sharp_young_rg}, we find
\begin{equation}
\begin{aligned}
\|r_{b,s}\|_\infty
&\le
\frac{b}{2\sqrt{s}}
\left(
\frac{b\sqrt{s(s-1)}}{bs-1}
\right)^{s-1}\\
&=\frac{\Gamma_{b,s}}{2}.
\end{aligned}
\label{eq:uniform_rg_density_bound}
\end{equation}
Consequently, for every $t\ge0$,
\[
\begin{aligned}
F_{\RG_\infty(U)}(t)
&=\int_{-t}^{t}r_{b,s}(x)\,dx\\
&\le\min\{1,\Gamma_{b,s}t\}\\
&=F_{\Gamma_{b,s}^{-1}U}(t).
\end{aligned}
\]
By the definition of the peakedness
order, this implies
\begin{equation}
\RG_\infty(U)
\succeq\Gamma_{b,s}^{-1}U.
\label{eq:uniform_rg_expansion}
\end{equation}

On the other hand, for a standard
Gaussian random variable $G$,
\[
F_G(t)
\le\min\left\{
1,2\phi(0)t
\right\}
\le\min\{1,t\}
=F_U(t),
\]
where we used $2\phi(0)=\sqrt{2/\pi}<1$.
Hence,
\[
G\succeq U.
\]
By Lemma~\ref{lem:order}, the
zero-temperature RG transformation
preserves the peakedness order.
Furthermore, it is positively
homogeneous:
\[
\RG_\infty(cY)
\overset{d}{=}c\RG_\infty(Y)
\qquad(c>0).
\]
Combining these properties with
Eq.~\eqref{eq:uniform_rg_expansion},
we obtain inductively
\begin{equation}
G_n=(\RG_\infty)^n(G)
\succeq\Gamma_{b,s}^{-n}U
\qquad(n\ge0).
\label{eq:iterated_uniform_comparison}
\end{equation}
Indeed, the assertion holds for
$n=0$ since $G\succeq U$.
If it holds for some $n\ge0$, then
\[
\begin{aligned}
G_{n+1}
&\overset{d}{=}\RG_\infty(G_n)\\
&\succeq
\RG_\infty(\Gamma_{b,s}^{-n}U)\\
&\overset{d}{=}
\Gamma_{b,s}^{-n}\RG_\infty(U)\\
&\succeq
\Gamma_{b,s}^{-(n+1)}U.
\end{aligned}
\]
Thus, Eq.~\eqref{eq:iterated_uniform_comparison}
holds for every integer $n\ge0$.
In particular, for every $t\ge0$,
\[
\begin{aligned}
F_{G_n}(t)
&\le
F_{\Gamma_{b,s}^{-n}U}(t)\\
&=\min\{1,\Gamma_{b,s}^nt\}.
\end{aligned}
\]
Since $g_n$ is symmetric and
continuous at the origin,
dividing by $2t$ and letting
$t\downarrow0$ yields
\[
g_n(0)\le\frac{\Gamma_{b,s}^n}{2}.
\]
Recalling that
\[
\alpha_n=\frac{\phi(0)}{g_n(0)},
\]
we conclude that
\[
\alpha_n
\ge2\phi(0)\Gamma_{b,s}^{-n}.
\]

Taking the $n$-th root and using
Lemma~\ref{Asymptotic growth rate},
we obtain
\[
\begin{aligned}
\Lambda(b,s)
&=\lim_{n\to\infty}\alpha_n^{1/n}\\
&\ge
\lim_{n\to\infty}
[2\phi(0)]^{1/n}\Gamma_{b,s}^{-1}\\
&=\Gamma_{b,s}^{-1}.
\end{aligned}
\]
Therefore, $\Gamma_{b,s}<1$
implies $\Lambda(b,s)>1$.

Finally, suppose that $d\ge3$, or equivalently,
$s\ge b^2$.
Since
\[
(bs-1)^2-b^2s(s-1)
=bs(b-2)+1>0,
\]
we have
\[
\Gamma_{b,s}
<
\frac{b}{\sqrt{s}}
\le1.
\]
Thus, $\Lambda(b,s)>1$ whenever $d\ge3$,
which completes the proof.
\end{proof}

\section{Discussion}

We have established a necessary and sufficient
criterion for the existence of finite-temperature
spin-glass order on MK hierarchical lattices
with arbitrary integers $b,s\ge2$,
in terms of the zero-temperature RG flow.
In particular, spin-glass order exists at
sufficiently low temperatures when
$\Lambda(b,s)>1$, whereas it is absent
at every finite temperature when
$\Lambda(b,s)\le1$.

Our results establish the existence of
finite-temperature spin-glass order for
$d\ge3$ and its absence for
\[
d<1+\frac{\log17}{\log6}\approx2.581.
\]
In the intermediate regime
\[
1+\frac{\log17}{\log6}\le d<3,
\]
the presence or absence of spin-glass
order is determined by $\Lambda(b,s)$
for each pair $(b,s)$.
Although our finite-step procedure
provides a classification in principle
whenever $\Lambda(b,s)\ne1$,
its computational cost increases with
$b$ and $s$, making its practical
implementation challenging for large
scale factors.
Developing simpler analytical criteria
for large $b$ and $s$ remains an
important open problem.

An important open problem is to determine
whether $\Lambda(b,s)=1$ can occur
for some integers $b,s\ge2$.
Resolving this question would clarify
whether the finite-step criterion
applies to every MK hierarchical lattice.

Our rigorous absence criterion may
appear to be in tension with previous
numerical estimates of the lower critical
dimension, which suggest a value of
approximately $2.52$~\cite{demirtacs2015lower}.
However, the previous study examined
a broader class of hierarchical lattices,
including structures beyond the MK construction.
In contrast, our rigorous analysis is restricted
to MK hierarchical lattices with arbitrary
integers $b,s\ge2$.
Therefore, our results are not necessarily
inconsistent with the previous numerical findings.

Our results establish the absence of
finite-temperature spin-glass order for
$d<3$ in the large-$b$ limit, while
guaranteeing its existence at sufficiently
low temperatures for all $d\ge3$.
Thus, $d=3$ emerges as the asymptotic
lower critical dimension of MK hierarchical
lattices in the large-$b$ limit, consistent with the prediction
of scaling theory~\cite{boettcher2005stiffness}.

%

Our proof of finite-temperature
spin-glass order exploits the
hierarchical structure of the MK
lattice to construct a reference
spin configuration, thereby enabling
a Peierls-type argument.
Whether this approach can be extended
to spin-glass models on conventional
$d$-dimensional lattices remains unclear.

Although the present analysis focuses
on Gaussian couplings, several key
ingredients of our approach may extend
to other symmetric, continuous, unimodal
coupling distributions.
Extensions to discrete coupling distributions
are also possible, although they require
additional technical modifications.
We plan to report these extensions elsewhere.

\subsection*{Note added}

After the completion of this work,
Pei~\cite{pei2026positive} independently
established the existence of finite-temperature
spin-glass order on MK hierarchical lattices
with $b=2$ and $s\ge4$.
For Gaussian couplings, the existence of
spin-glass order was also proved for
$(b,s)=(2,3)$ using computer-assisted
rigorous estimates.

Pei's proof directly controls two-replica
overlaps along hierarchical ancestor chains,
without introducing an explicit reference
spin configuration.
In contrast, our approach employs a
Peierls-type argument based on a reference
configuration and provides a unified criterion
for the existence and absence of spin-glass
order for arbitrary $b,s\ge2$.

\appendix
\section*{Data availability statement}
No new data were created or analyzed in this study.

\section*{Acknowledgment}
This work was supported by JST BOOST, Japan Grant Number JPMJBY24B6.
This work was also supported by JSPS KAKENHI Grant No. 24K16973.
In addition, this work was supported by the Cross-ministerial Strategic Innovation Promotion Program (SIP) of the Cabinet Office (No. 23836436).

\section{Proof of Lemma~\ref{lem:dimension_bound}}
\label{app:gamma_bounds}

We establish the bounds on $\gamma_b$ used
in the proof of Lemma~\ref{lem:dimension_bound}.
Recall from Eq.~\eqref{eq:gamma} that
\[
\gamma_b
=
\mathbb E\left[\min_{1\le j\le b}Z_j^2\right]
=
2\int_0^\infty tS(t)^b\,dt,
\]
where
\[
S(t)=\operatorname{erfc}(t/\sqrt2)
\]
is the survival function of the absolute
value of a standard Gaussian random variable.

We first derive the bound
$\gamma_b\le6/[(b+1)(b+2)]$.
Let
\[
F(t):=\mathbb P(|Z|\le t),
\qquad \xi:=F^{-1}.
\]
Since $F$ is continuous and strictly
increasing on $[0,\infty)$, the probability
integral transform gives
\[
U_j:=F(|Z_j|)\sim\mathrm{Unif}(0,1),
\qquad j=1,\ldots,b,
\]
where the $U_j$ are independent.
Consequently,
\[
|Z_j|=\xi(U_j)
\]
almost surely.
Since $\xi$ is nonnegative and increasing,
\[
\min_{1\le j\le b}Z_j^2
=
\xi\left(\min_{1\le j\le b}U_j\right)^2.
\]
The minimum
\[
U_{(1)}:=\min_{1\le j\le b}U_j
\]
has probability density
\[
f_{U_{(1)}}(u)=b(1-u)^{b-1},
\qquad 0<u<1.
\]
Therefore,
\begin{equation}
\gamma_b
=
b\int_0^1\xi(u)^2(1-u)^{b-1}\,du.
\label{eq:gamma_quantile}
\end{equation}
Since $F$ is concave on $[0,\infty)$,
its inverse $\xi$ is convex and satisfies
$\xi(0)=0$.
Hence,
\[
R(u):=\frac{\xi(u)^2}{u^2}
\]
is nondecreasing on $(0,1)$.
Define
\[
w_b(u):=
\frac{b(b+1)(b+2)}{2}
u^2(1-u)^{b-1}.
\]
Then $w_b$ is a probability density,
and Eq.~\eqref{eq:gamma_quantile} becomes
\[
(b+1)(b+2)\gamma_b
=
2\int_0^1R(u)w_b(u)\,du.
\]
Since
\[
\frac{w_{b+1}(u)}{w_b(u)}
=
\frac{b+3}{b}(1-u)
\]
is decreasing in $u$, the distribution
with density $w_{b+1}$ is stochastically
smaller than that with density $w_b$.
As $R$ is nondecreasing, we obtain
\begin{equation}
(b+2)(b+3)\gamma_{b+1}
\le
(b+1)(b+2)\gamma_b.
\label{eq:gamma_monotonicity}
\end{equation}
Since $\gamma_1=1$, iteration yields
\[
\gamma_b\le
\frac{6}{(b+1)(b+2)}.
\]

We next establish the bound
$\gamma_b\le\pi/b^2$.
The Gaussian Mills inequality~\cite{gordon1941values} gives
\[
S(t)\le
\sqrt{\frac2\pi}
\frac{e^{-t^2/2}}{t},
\qquad t>0.
\]
Consequently, the hazard rate
\[
h(t):=-\frac{S'(t)}{S(t)}
\]
satisfies
\[
h(t)\ge t,
\qquad
h'(t)=h(t)[h(t)-t]\ge0.
\]
Thus, $\log S(t)$ is concave.
Since
\[
S(0)=1,
\qquad
S'(0)=-\sqrt{\frac2\pi},
\]
concavity implies
\[
S(t)\le
\exp\left(-\sqrt{\frac2\pi}\,t\right).
\]
Substituting this bound into the
integral representation of $\gamma_b$,
we obtain
\[
\begin{aligned}
\gamma_b
&=2\int_0^\infty tS(t)^b\,dt\\
&\le
2\int_0^\infty
t\exp\left(
-b\sqrt{\frac2\pi}\,t
\right)dt\\
&=\frac{\pi}{b^2}.
\end{aligned}
\]
This proves
Eq.~\eqref{eq:gamma_general_bounds}.

\subsection*{Explicit bounds for small $b$}

We next establish the bounds required
for $2\le b\le6$.

For $b=2$, a direct calculation gives
\[
\gamma_2=1-\frac2\pi<\frac12.
\]

For $3\le b\le6$, we derive an explicit
upper bound on the Gaussian survival
function.
Set
\[
P(t):=
t-\frac{t^3}{6}
+\frac{t^5}{40}
-\frac{t^7}{336},
\qquad
c_*:=\frac{7977}{10000}.
\]
The Taylor expansion of $e^{-t^2/2}$
with a remainder of known sign gives
\[
\int_0^t e^{-u^2/2}\,du
\ge P(t)
\qquad(0\le t\le2).
\]
Since $P(t)\ge0$ on $[0,2]$ and
$c_*<\sqrt{2/\pi}$, we obtain
\begin{equation}
S(t)\le1-c_*P(t)
\qquad(0\le t\le2).
\label{eq:survival_polynomial}
\end{equation}
For $t\ge2$, the Gaussian Mills inequality
implies $h(t)\ge t\ge2$ and hence
\[
S(t)\le S(2)e^{-2(t-2)}.
\]
Moreover,
\[
S(2)
\le\sqrt{\frac2\pi}\frac{e^{-2}}2
<\frac3{50}.
\]
Therefore,
\begin{align}
\gamma_b
&=2\int_0^\infty tS(t)^b\,dt
\nonumber\\
&\le
2\int_0^2
t[1-c_*P(t)]^b\,dt
+2\left(\frac3{50}\right)^b
\left(\frac1b+\frac1{4b^2}\right)
\nonumber\\
&=:H_b.
\label{eq:gamma_H_bound}
\end{align}
The integral defining $H_b$ is a
polynomial integral with rational
coefficients and can therefore be
evaluated exactly.
Indeed, writing
\[
[1-c_*P(t)]^b
=
\sum_{r=0}^{7b}a_{b,r}t^r,
\]
where all $a_{b,r}$ are rational,
we have
\[
\begin{aligned}
H_b
&=
\sum_{r=0}^{7b}
\frac{2^{r+3}}{r+2}a_{b,r}
+
2\left(\frac3{50}\right)^b
\left(\frac1b+\frac1{4b^2}\right).
\end{aligned}
\]
Evaluating these finite sums using exact
rational arithmetic yields
\begin{align}
H_3&<\frac{49}{250}<\frac15,
\nonumber\\
H_4&<\frac{61}{500}<\frac18,
\nonumber\\
H_5&<\frac{52}{625}<\frac1{12},
\nonumber\\
H_6&<\frac{61}{1000}<\frac3{49}.
\label{eq:gamma_six}
\end{align}
Since $\gamma_b\le H_b$, these
inequalities establish the required
bounds for $3\le b\le6$.

\subsection*{Refined bound for $b\ge6$}

Finally, combining
$\gamma_6<3/49$ with the monotonicity
in Eq.~\eqref{eq:gamma_monotonicity},
we obtain
\[
\begin{aligned}
\gamma_b
&\le
\frac{56\gamma_6}{(b+1)(b+2)}\\
&<
\frac{24}{7(b+1)(b+2)}
\end{aligned}
\]
for every $b\ge6$.
This proves
Eq.~\eqref{eq:gamma_refined_bound}.
In particular,
\[
\gamma_7<\frac1{21},
\qquad
\gamma_8<\frac4{105},
\qquad
\gamma_9<\frac{12}{385}.
\]
Together with the bounds established
above for $2\le b\le6$, these inequalities
complete the finite-case verification
used in the proof of
Lemma~\ref{lem:dimension_bound}.

\section{Proof of Lemma~\ref{lem:finite_zero_convergence}}
\label{app:finite_zero_density}

Throughout this appendix, we assume $b,s\ge2$.

For each $r\ge0$ and finite $\beta\ge1$,
let $q_{r,\beta}$ denote the density of
the series output obtained from $b$
independent copies of $J_r(\beta)$.
Similarly, let $q_{r,\infty}$ denote
the density of the zero-temperature
series output obtained from $b$
independent copies of $G_r$.
Thus,
\[
p_{r+1,\beta}=q_{r,\beta}^{*s},
\qquad
g_{r+1}=q_{r,\infty}^{*s}.
\]
By Lemma~\ref{preservation-unimodality},
the densities $p_{r,\beta}$, $g_r$,
$q_{r,\beta}$, and $q_{r,\infty}$
are symmetric unimodal.
We choose even representatives that
are nonincreasing on $(0,\infty)$.

\subsection*{A uniform comparison of the two RG maps}

For every finite $\beta>0$ and
$x_1,\ldots,x_b\in\mathbb R$,
\begin{equation}
\left|
\Phi_\beta(x_1,\ldots,x_b)
-
\Phi_\infty(x_1,\ldots,x_b)
\right|
\le\frac{\log b}{2\beta}.
\label{eq:series_zero_error}
\end{equation}
To prove this, observe that the
finite- and zero-temperature series
outputs have the same sign.
For nonnegative inputs $u,v$,
the two-bond series transformation satisfies
\[
e^{-2\beta\Phi_\beta(u,v)}
=
\frac{e^{-2\beta u}+e^{-2\beta v}}
{1+e^{-2\beta(u+v)}}
\le
e^{-2\beta u}+e^{-2\beta v}.
\]
By associativity of the series
transformation, this implies
\[
\begin{aligned}
e^{-2\beta|\Phi_\beta(x_1,\ldots,x_b)|}
&\le
\sum_{j=1}^b e^{-2\beta|x_j|}\\
&\le
b e^{-2\beta\min_j|x_j|}.
\end{aligned}
\]
Since
\[
|\Phi_\beta(x_1,\ldots,x_b)|
\le\min_j|x_j|,
\]
we obtain
\[
0\le
\min_j|x_j|
-
|\Phi_\beta(x_1,\ldots,x_b)|
\le\frac{\log b}{2\beta}.
\]
Together with the equality of the
output signs, this proves
Eq.~\eqref{eq:series_zero_error}.

The zero-temperature series transformation
also satisfies
\begin{equation}
\begin{aligned}
&\left|
\Phi_\infty(x_1,\ldots,x_b)
-
\Phi_\infty(y_1,\ldots,y_b)
\right|\\
&\qquad\le
\max_{1\le j\le b}|x_j-y_j|.
\end{aligned}
\label{eq:signed_min_lipschitz}
\end{equation}
Indeed, if the two outputs have the
same sign, this follows from the
Lipschitz property of the minimum
of the input magnitudes.
If they have opposite signs,
at least one input coordinate
changes sign, and the change in
that coordinate is at least the
sum of the two output magnitudes.
The cases involving a zero output
follow by continuity.

We now construct the finite- and
zero-temperature RG flows using
the same microscopic Gaussian couplings.
Suppose that the effective couplings
at generation $r$ differ by at most
$e_r$ in this coupling.
By
Eqs.~\eqref{eq:series_zero_error}
and~\eqref{eq:signed_min_lipschitz},
the error after the next RG step
is bounded by
\[
e_{r+1}
\le
s e_r+\frac{s\log b}{2\beta}.
\]
Since $e_0=0$, iteration gives
\[
e_r
\le
\frac{s\log b}{2\beta}
\frac{s^r-1}{s-1}.
\]
Consequently,
\begin{equation}
|J_r(\beta)-G_r|
\le
\frac{s\log b}{2\beta}
\frac{s^r-1}{s-1}
=:\varepsilon_r(\beta)
\label{eq:appendix_coupling_error}
\end{equation}
almost surely under the common-input
coupling.

In particular, for every fixed $r$,
the finite-temperature RG output
converges almost surely to its
zero-temperature counterpart as
$\beta\to\infty$.
The same conclusion holds for
the corresponding series outputs.

\subsection*{Uniform density estimates}

We next prove by induction that,
for every fixed $r\ge0$,
\begin{equation}
\sup_{\beta\ge1}
\|p_{r,\beta}\|_\infty<\infty,
\qquad
\|g_r\|_\infty<\infty.
\label{eq:appendix_gaussian_uniform_density}
\end{equation}
For $r=0$, both assertions follow from
\[
p_{0,\beta}=g_0=\phi.
\]

We first establish a lower bound
on the magnitude of the series output.
For nonnegative $u,v$,
\[
\Phi_\beta(u,v)
=
\frac{1}{2\beta}
\log
\frac{\cosh(\beta(u+v))}
{\cosh(\beta(u-v))}.
\]
Define
\[
\chi(z):=
z\tanh z-\log\cosh z,
\qquad z\ge0.
\]
Since
\[
\chi'(z)=z\operatorname{sech}^2z\ge0,
\]
differentiation yields
\[
\frac{\partial}{\partial\beta}
\Phi_\beta(u,v)
=
\frac{
\chi(\beta(u+v))
-\chi(\beta|u-v|)
}{2\beta^2}
\ge0.
\]
By associativity and coordinatewise
monotonicity of the series transformation
for nonnegative inputs, this property
extends to any number of bonds.
Thus, for every $\beta\ge1$,
\[
|\Phi_\beta(x_1,\ldots,x_b)|
\ge
|\Phi_1(x_1,\ldots,x_b)|.
\]
The same inequality holds for
the zero-temperature transformation.

Furthermore, using
$\operatorname{arctanh}z\ge z$
for $0\le z<1$, we obtain
\[
\begin{aligned}
|\Phi_1(x_1,\ldots,x_b)|
&\ge
\prod_{j=1}^b\tanh|x_j|\\
&\ge
c_b\prod_{j=1}^b\min\{|x_j|,1\},
\end{aligned}
\]
where
\[
c_b:=(\tanh1)^b>0.
\]
Here, we used the concavity of
$\tanh$ on $[0,1]$ and its
monotonicity on $[0,\infty)$.
Consequently,
\begin{equation}
|\Phi_\beta(x_1,\ldots,x_b)|
\ge
c_b\prod_{j=1}^b\min\{|x_j|,1\}
\label{eq:unbounded_series_lower}
\end{equation}
for every $\beta\ge1$,
and also for $\beta=\infty$.

Suppose that
Eq.~\eqref{eq:appendix_gaussian_uniform_density}
holds at generation $r$.
Set
\[
M:=
\max\left\{
\sup_{\beta\ge1}\|p_{r,\beta}\|_\infty,
\|g_r\|_\infty
\right\}<\infty.
\]

Consider $b$ independent random
variables $Y_1,\ldots,Y_b$,
each having either the density
$p_{r,\beta}$ for some $\beta\ge1$
or the density $g_r$.
In either case, the common input
density is symmetric and bounded by $M$.

Introduce the auxiliary variables
\[
U_j:=
-\log\min\{|Y_j|,1\},
\qquad 1\le j\le b.
\]
Each $U_j$ is nonnegative and may
have an atom at zero.
On $(0,\infty)$, its density is
bounded by
\[
2e^{-u}M.
\]
Expanding the convolution of these
distributions into their atomic
and absolutely continuous parts,
we obtain, for every $L>0$,
\begin{equation}
\begin{aligned}
&\mathbb P\left(
\sum_{j=1}^bU_j\ge L
\right)\\
&\quad\le
e^{-L}
\sum_{j=1}^b
\binom bj(2M)^j
\sum_{i=0}^{j-1}\frac{L^i}{i!}.
\end{aligned}
\label{eq:log_product_bound}
\end{equation}
Indeed, a contribution involving
$j$ absolutely continuous factors
is bounded by the density
\[
(2M)^j e^{-u}
\frac{u^{j-1}}{(j-1)!},
\qquad u>0,
\]
whose integral over $[L,\infty)$ is
\[
(2M)^j e^{-L}
\sum_{i=0}^{j-1}\frac{L^i}{i!}.
\]
The remaining factors contribute
atoms at zero, whose masses are
at most one.

By Eq.~\eqref{eq:unbounded_series_lower},
for $0<t<c_b$,
\[
\begin{aligned}
&\{|\Phi_\beta(Y_1,\ldots,Y_b)|\le t\}\\
&\quad\subseteq
\left\{
\sum_{j=1}^bU_j
\ge\log(c_b/t)
\right\}.
\end{aligned}
\]
The same inclusion holds with
$\Phi_\infty$ in place of $\Phi_\beta$.

It follows from
Eq.~\eqref{eq:log_product_bound}
that there exist constants
$A_r<\infty$ and $t_r\in(0,1)$,
independent of $\beta$, such that
\[
\mathbb P\left(
|\Phi_\beta(Y_1,\ldots,Y_b)|\le t
\right)
\le
A_rt(1+|\log t|^{b-1})
\]
for every $0<t<t_r$.
The same estimate holds for
the zero-temperature series output.

By Lemma~\ref{preservation-unimodality},
the series-output densities
$q_{r,\beta}$ and $q_{r,\infty}$
are symmetric and nonincreasing
on the positive half-line.
Thus, for $x>0$,
\[
2xq_{r,\beta}(x)
\le
2\int_0^xq_{r,\beta}(y)\,dy,
\]
and the analogous inequality
holds for $q_{r,\infty}$.

Combining these inequalities with
the preceding small-ball estimates,
and increasing $A_r$ if necessary,
we obtain the common upper bound
\begin{equation}
q_{r,\beta}(x),\ q_{r,\infty}(x)
\le
\begin{cases}
A_r(1+|\log|x||^{b-1}),
&0<|x|<t_r,\\
(2|x|)^{-1},
&|x|\ge t_r.
\end{cases}
\label{eq:unbounded_density_envelope}
\end{equation}
Here, the second bound follows
from the normalization and symmetric
unimodality of the densities.

The right-hand side belongs to
$L^2(\mathbb R)$.
Consequently,
\begin{equation}
\sup_{\beta\ge1}\|q_{r,\beta}\|_2<\infty,
\qquad
\|q_{r,\infty}\|_2<\infty.
\label{eq:appendix_uniform_ltwo}
\end{equation}

Since $s\ge2$ and all branch-output
densities have $L^1$ norm one,
Young's convolution inequality gives
\[
\begin{aligned}
\|p_{r+1,\beta}\|_\infty
&=
\|q_{r,\beta}^{*s}\|_\infty\\
&\le
\|q_{r,\beta}^{*2}\|_\infty\\
&\le
\|q_{r,\beta}\|_2^2.
\end{aligned}
\]
Similarly,
\[
\begin{aligned}
\|g_{r+1}\|_\infty
&=
\|q_{r,\infty}^{*s}\|_\infty\\
&\le
\|q_{r,\infty}\|_2^2.
\end{aligned}
\]

Both right-hand sides are finite,
and the first is uniformly bounded
in $\beta\ge1$.
This completes the induction and
establishes
Eq.~\eqref{eq:appendix_gaussian_uniform_density}.

We emphasize that no bounded-support
assumption has been imposed on the
input distributions.

\subsection*{Convergence of the densities}

We now prove that, for every fixed
$r\ge0$,
\begin{equation}
\|q_{r,\beta}-q_{r,\infty}\|_2
\longrightarrow0
\qquad(\beta\to\infty).
\label{eq:appendix_ltwo_convergence}
\end{equation}

The common-input comparison
established above implies that
the finite-temperature series outputs
converge almost surely to their
zero-temperature counterparts.
Their distributions therefore converge
weakly to the distribution with
density $q_{r,\infty}$.

For $x>0$ and $0<h<x$,
symmetric unimodality gives
\[
\begin{aligned}
\frac1h\int_x^{x+h}
q_{r,\beta}(u)\,du
&\le q_{r,\beta}(x)\\
&\le
\frac1h\int_{x-h}^{x}
q_{r,\beta}(u)\,du.
\end{aligned}
\]
Since the limiting distribution
is absolutely continuous,
weak convergence implies convergence
of both interval probabilities.
Letting $\beta\to\infty$ and then
$h\downarrow0$, we obtain
\[
q_{r,\beta}(x)
\longrightarrow q_{r,\infty}(x)
\]
at every positive continuity point
of $q_{r,\infty}$.

Since a monotone function has at
most countably many discontinuities,
and the densities are symmetric,
the convergence holds for almost
every $x\in\mathbb R$.

By Eq.~\eqref{eq:unbounded_density_envelope},
both $q_{r,\beta}$ and $q_{r,\infty}$
are bounded by a common function
in $L^2(\mathbb R)$.
The dominated convergence theorem
therefore yields
\[
\lim_{\beta\to\infty}
\|q_{r,\beta}-q_{r,\infty}\|_2=0,
\]
which proves
Eq.~\eqref{eq:appendix_ltwo_convergence}.

Finally, using
\[
p_{r+1,\beta}=q_{r,\beta}^{*s},
\qquad
g_{r+1}=q_{r,\infty}^{*s},
\]
we expand the difference of the
two convolution powers by telescoping.

In each term, we place the factor
$q_{r,\beta}-q_{r,\infty}$
and one additional branch density
in $L^2(\mathbb R)$.
The remaining $s-2$ factors have
$L^1$ norm one.
By Young's convolution inequality,
\[
\begin{aligned}
&\|p_{r+1,\beta}-g_{r+1}\|_\infty\\
&\quad\le
s\max\left\{
\|q_{r,\beta}\|_2,
\|q_{r,\infty}\|_2
\right\}\\
&\qquad\times
\|q_{r,\beta}-q_{r,\infty}\|_2.
\end{aligned}
\]
By
Eqs.~\eqref{eq:appendix_uniform_ltwo}
and~\eqref{eq:appendix_ltwo_convergence},
the right-hand side tends to zero
as $\beta\to\infty$.
Hence,
\[
\lim_{\beta\to\infty}
\|p_{r+1,\beta}-g_{r+1}\|_\infty=0.
\]

For $r=0$, the initial densities coincide:
\[
p_{0,\beta}=g_0=\phi.
\]
Therefore,
\[
\lim_{\beta\to\infty}
\|p_{r,\beta}-g_r\|_\infty=0
\]
for every fixed integer $r\ge0$.

This proves
Eq.~\eqref{eq:finite_zero_density}
and completes the proof of
Lemma~\ref{lem:finite_zero_convergence}.

\bibliographystyle{apsrev4-1}
\bibliography{references}

%
%
%
%
%
%
%
%
%
%

\end{document}